\documentclass{aa}  
\usepackage{graphicx}
\usepackage{txfonts}

\usepackage{lscape}
\usepackage{amssymb}
\usepackage{natbib}  
\bibpunct{(}{)}{;}{a}{,}{,} 
\usepackage{longtable}
\usepackage{graphicx}
\usepackage[hang,bf,footnotesize]{subfigure}
\usepackage[figuresright]{rotating}

\begin{document}
%5%%%%%%%%%%%%%%%%%%%%%%%%%%%%%%%%%%%%%%%%%%%%%%%%%%%%%%%%%%%%%%%%%%%%%
\titlerunning{Disks frequency and age in M16}	
   \title{Chronology of star formation and disk evolution in the Eagle Nebula}

%    \subtitle{Disk frequency and sequential star formation in the Eagle Nebula}

   \author{M. G. Guarcello\inst{1,2,3} \and G. Micela\inst{2} \and G. Peres\inst{1} \and L. Prisinzano\inst{2}     \and S. Sciortino\inst{2}}

   \offprints{mguarce@astropa.unipa.it}

   \institute{Dipartimento di Scienze Fisiche ed Astronomiche, Universit\'a di Palermo, Piazza del Parlamento 1, I-90134 Palermo Italy
   \and
   INAF - Osservatorio Astronomico di Palermo, Piazza del Parlamento 1, 90134 Palermo Italy \and
   Actually at Smithsonian Astrophysical Observatory, MS-3, 60 Garden Street, Cambridge, MA 02138, USA} 

  \date{}

% 5 {} token are mandatory
 
  \abstract
   {Massive star-forming regions are characterized by intense ionizing fluxes, strong stellar winds and, occasionally, supernovae explosions, all of which have important effects on the surrounding media, on the star-formation process and on the evolution of young stars and their circumstellar disks. We present a multiband study of the massive young cluster NGC~6611 and its parental cloud (the Eagle Nebula) with the aim of studying how OB stars affect the early stellar evolution and the formation of other stars.}
   {We search for evidence of: triggering of star formation by the massive stars inside NGC~6611 on a large spatial scale ($\sim 10$ parsec) and ongoing disk photoevaporation in NGC~6611 and how its efficiency depends on the mass of the central stars.}
   {We assemble a multiband catalog of the Eagle Nebula with photometric data, ranging from B band to 8.0 $\mu m$, and X-ray data obtained with two new and one archival CHANDRA/ACIS-I observation. We select the stars with disks from infrared photometry and disk-less ones from X-ray emission, which are associated both with NGC~6611 and the outer region of the Eagle Nebula. We study induced photoevaporation searching for the spatial variation of disk frequency for distinct stellar mass ranges. The triggering of star formation by OB stars has been investigated by deriving the history of star formation across the nebula.}
   {We find evidence of sequential star formation in the Eagle Nebula going from the southeast (2.6 Myears) to the northwest (0.3 Myears), with the median age of NGC~6611 members $\sim$1 Myear. In NGC~6611, we observe a drop of the disk frequency close to massive stars (up to an average distance of 1 parsec), without observable effects at larger distances. Furthermore, disks are more frequent around low-mass stars ($\leq 1 M_{\odot}$) than around high-mass stars, regardless of from the distance from OB stars.}
   {The star formation chronology we find in the Eagle Nebula does not support the hypothesis of a large-scale process triggered by OB stars in NGC~6611. Instead, we speculate that it was triggered by the encounter (about 3 Myears ago) with a giant molecular shell created by supernovae explosions about 6 Myears ago. We find evidence of disk photoevaporation close to OB stars, where disks are heated by incident extreme ultraviolet (EUV) radiation. No effects are observed at large distances from OB stars, where photoevaporation is induced by the far ultraviolet (FUV) radiation, and long timescales are usually required to completely dissipate the disks.}


   \maketitle
%
%________________________________________________________________

\section{Introduction}
\label{intro}

	In the first decade of the 21th century, a large number of active star-forming sites in our Galaxy have been identified and studied in detail with a multi-wavelength approach, allowing to advance our understanding of star formation (see, for example, \citealt{Feige99}), even if we yet lack a full understanding of how molecular clouds turn into stellar clusters. In particular, it is still unclear what initiates the cloud collapse and what the environmental effects are on the stellar population formed after the collapse. \par
The initial impulse for cloud collapse may be provided by shocks created by intense stellar winds from a large amount of massive stars or supernovae explosions \citep{Cho07}. Sometimes they can form giant molecular shells, which can be associated to star-formation events, such as the Barnard loop and the Eridanus loop in the Orion complex \citep{Rey79}. \par
	It is a debated question whether the stellar population produced by the collapse of a molecular cloud depends on the environment in which stars form. This question is also related to the universal relation between the cluster age and the fraction of members that still bear a disk (i.e. the disk frequency, \citealp{Hai01}). In particular, the role played by OB stars in young clusters is unclear. Their energetic radiation, which ionizes the HI clouds into HII, and their intense winds disperse the cloud, which halts the star-formation process. On the other hand, the high pressure provided by their stellar wind and energetic radiation may compress the clouds, inducing their radiative implosion and the formation of subsequent generations of stars \citep{Mck07}. The intense UV radiation from massive stars may also dissipate the protostellar cores emerged from the evaporating parental clouds \citep{Heste96} and the nearby circumstellar disks by photoevaporation (i.e. \citealp{Sto99}), which affects the mass accretion onto protostars and the formation of planetary systems. \par
The effects of massive stars on the evolution of Young Stellar Objects (YSOs) and on the star-formation process can be analyzed with several methods. For instance, for young, not yet relaxed clusters, the study of the sequential star formation from the positions of massive stars outward can give a clue about the relevance of induced cloud-compression on the formation of new generations of stars. This approach has been successfully applied in several cases, such as the HII regions RCW~79 \citep{Zava06}, Sh~217, Sh~219 \citep{Deha03}, as well as in the star-forming regions 30~Doradus \citep{Wal97} and NGC~1893 \citep{Negu07}. \par
	The outcomes of the investigation on disk frequencies and Initial Mass Function (IMF) in young clusters are much more debated. It is well known that the disk frequency decreases on average with increasing cluster age \citep{Hai01}. There is no general consensus yet whether the observed disk frequencies are independent of the cluster population of OB stars (as in NGC~6611, \citealt{Oli05}) or if instead they decrease close to OB stars as a consequence of externally induced photoevaporation (as in NGC~2244, \citealt{Balo07}). Furthermore, the universality of the IMF seems to imply that the dissipation of protostellar cores by the UV radiation from massive stars does not significantly affect the mass accretion onto YSOs. \par
We investigate these open issues for the Eagle Nebula (M16) and the associated young cluster NGC~6611. In Sect. \ref{m16} we review some of the main characteristics of both the Eagle Nebula and NGC~6611, mostly focusing on the morphology of the region and on the existing evidence about the feedback by massive stars on the star formation and disk evolution. Since a multi-wavelengths approach is necessary to select and classify T-Tauri stars and to obtain information about their stellar parameters, in this work we analyze a set of optical, infrared, and X-rays data, which are described in Sect. \ref{datasec}. These data are combined in in a multi-band catalog of the region, which is described in Sect. \ref{catsec}. Appendix \ref{matchapp} is devoted to the procedure used to match all the data and obtain the multi-band catalog. The selection of stars with infrared excesses due to a circumstellar disk, performed with suitable reddening-free color indices, optical and infrared color-magnitude diagrams and the IRAC color-color diagram is described in Sect. \ref{disksec}. In that section we also analyze the contamination of the list of selected candidate stars associated to M16 by both foreground and background objects. The final member list is summarized in Sect. \ref{listsec}. The effects of the UV radiation that is emitted by massive stars on the evolution of circumstellar disks in M16 are analyzed in Sect. \ref{diskfresec}, where we also compare the timescales for externally induced disk photoevaporation and ``normal'' dissipation at different masses of the central star. The triggering of star formation in M16 by the massive stars in NGC~6611 is analyzed in Sect. \ref{agesec}, where we find finding the chronology of star formation across the cloud. All the results are discussed in detail in Sect. \ref{discsec}. In Appendix \ref{popusec} we describe the stellar content of three peculiar regions of the Eagle Nebula; in Appendix \ref{extapp} we derive the extinction map of the cloud; while in Appendix \ref{matchapp} we describe the adopted catalog cross-matching procedure.
%%%%%%%%%%%%%%%%%%%%%%%%%%%%%%%%%%%%%%%%%%%%%%%%%%%%%%%%%%%%%%%%%%%%%%%%%%%%%%%%%%%%%%%%%%%%%%%%%%%%%%%%%%%%%%%%%%%%%%%%%%%%%%%%%%%%%%

\section{The Eagle Nebula (M16) and NGC~6611}
\label{m16}

The Eagle Nebula, one of the most active star-forming sites in the Sagittarius arm, and the young open cluster NGC~6611 in its center are ideal targets to study star formation in the presence of massive stars. The distance to the cloud has been evaluated by several authors; here, we adopt the value computed in \citet{Io07}, hereafter GPM07, which is equal to 1750 parsec. \par
	Starting from the study of \citet{Dewi97}, several works have been devoted to the massive stars of NGC~6611. The cluster contains a remarkable population of OB stars (54, with 13 O stars, \citealp{Hille93}), which are concentrated mostly in the central cavity of the molecular cloud, which is about $5.5^{\prime}$ (2.2 parsec) wide. These massive stars are responsible for an intense UV field in the cluster center; for comparison, the three most massive stars in NGC~6611 alone emit about five times the ionizing flux of the O stars in the center of the Trapezium in the Orion Nebula Cluster (ONC), where disk photoevaporation has been directly observed \citep{Sto99}. The most massive star in the cluster is W205 ($\alpha =$18$^h$:18$^m$:36$^s$.48, $\delta =$~-13$^d$:48$^m$:02$^s$.5) with a mass of 75-80$M_{\odot}$ and a spectral class O3-O5V \citep{Hille93,Eva05}. The main source of ionizing flux in the nebula is W205, accounting for half of the total UV flux \citep{Heste96}.\par
Only few OB stars are far away from the cluster center. The most noticeable among them is W584, an O9V star at $\alpha =$18:18:23.69, $\delta =$-13:36:28.1, about $6.5^{\prime}$ from the approximate cluster center in the northwest direction. As inferred by the [8.0] IRAC image of the Eagle Nebula (\citealt{Io09}, hereafter GMD09), this star should have created a cavity in the nebula of about $1.5^{\prime}$ of radius (0.7 parsec). \par
	Several studies of the Eagle Nebula concern the triggering of star formation by the massive members and their influence on the evolution of both circumstellar disks and YSOs. \citet{Hille93} found that the massive population of NGC~6611 has a mean age of 2 Myears, with a spread of 1 Myear, even if one of the massive members (with a mass of 30$M_{\odot}$) is about 6 Myear old. Subsequent studies in the central region of NGC~6611 discovered a large age spread also for the low-mass members, whose age ranges from less than 1 Myear to $\sim 3$ Myears (GPM09, \citealt{Inde07}). The reasons for this age spread and the sequence of star formation are still unclear, but \citet{Gvara08} hypothesized that the oldest massive stars are blue stragglers, while the younger pre-main sequence (PMS) stars are part of distinct generations formed in successive star-formation episodes. It is reasonable to assume that the oldest massive stars triggered the star formation in the central region of NGC~6611 in the last 2 Myears. However, both \citet{Beli00} and \citet{Inde07} found no clue for triggered star-formation in the central region of the cluster. \par
Even if no proof of sequential star formation has been found in NGC~6611, there are indications of a short spatial scale feedback of OB stars on the star formation and the evolution of the cloud. For instance, the radiation from OB stars continuously models some cloud structures, such as the well known ``Pillars of Creation'', which contain deeply embedded disk-bearing stars \citep{Heste96,Mc02,Hea04,Sugi07}. \par
	Other active star-forming sites are present in the Eagle Nebula. The Bright Rimmed Cloud SFO30 is a region bright in IRAC bands, $\sim6^{\prime}$ away from the cluster center ($\sim3$ parsec) northward, where water maser and sources with excesses in near-infrared (NIR) bands have been found (\citealt{Hille93,Hea04,Inde07}, GMD09). An embedded massive star in formation has been found $8^{\prime}$ (3.8 parsec) west of the cluster center \citep{Inde07}. In the opposite direction ($\sim7^{\prime}$ eastward from the cluster center, 3.3 parsec) there is another elongated structure called ColumnV, probably modeled by the radiation from the cluster massive stars, where star-formation activity is proved by water masers \citep{Hea04} and candidate Herbig-Haro objects \citep{Mea86}. Finally, \citet{Inde07} found a new cluster of embedded objects inside a dense region of the Eagle Nebula, $16^{\prime}$ northeast of the cluster center. \citet{Io09} found that due to the strong absorption, only one of these young stars has a faint optical counterpart. Some of these active star-formation sites will be discussed in more detail in Appendix \ref{popusec}. \par
The effects of massive stars on star formation and disk evolution in NGC~6611 have been studied by other authors: \citet{Oli05} evaluated the disk frequency in the central region of the cluster to be equal to 58\%, comparable with that found in clusters with smaller populations of massive stars; \citet{Oli09} obtained the IMF of candidate cluster members, which has the same slope of the IMF as that of the Taurus region. These studies suggest that in NGC~6611 massive stars have no evident effect on the evolution of the other cluster members. On the other hand, GPM07 and GMD09 found that the disk frequency in the central region of the cluster decreases at small distance from massive stars, with a minimum value of $\sim16\%$ in the regions where the disks are illuminated by intense UV flux. Their results are compatible with a scenario in which UV radiation from massive stars destroyed the nearby circumstellar disks by induced photoevaporation. \par 	
	\citealt{Io09} compiled a list of 1264 stars associated with M16: 474 ClassI/II objects selected for their infrared excesses in 2MASS and IRAC bands in the $33^{\prime}\times 34^{\prime}$ field centered on NGC~6611 and 790 ClassIII PMS stars selected for their X-ray emission in the $17^{\prime}\times 17^{\prime}$ region centered on NGC~6611. In this paper we extend the selection of disk-less members to the entire Eagle Nebula region. Catalogs of the whole region have been published by \citealt{Hille93} (1026 sources with BVIJHK photometry), \citealt{Beli99} (2200 sources with BVJHK photometry) and GPM07 (38995 with BVIJHK and X-ray data). \par
%%%%%%%%%%%%%%%%%%%%%%%%%%%%%%%%%%%%%%%%%%%%%%%%%%%%%%%%%%%%%%%%%%%%%%%%%%%%%%%%%%%%%%%%%%%%%%%%%%%%%%%%%%%%%%%%%%%%%%%%%%%%%%%%%%%%%%

\section{Data analysis}
\label{datasec}

We compiled an optical-infrared-X-rays catalog of sources in the $33^{\prime} \times 34^{\prime}$ region centered on NGC~6611. The main improvements with respect to the catalog published by GPM07 are the use of new X-ray data in the outer regions of the nebula and the extension to fainter magnitudes in the infrared $JHK$ bands, taking advantage of the UKIDSS/GPS survey (see Sect. \ref{ukidssec}). Moreover, the instrument used in the UKIDSS surveys has an improved spatial resolution compared to that of the 2MASS survey. This is crucial in studies of crowded regions such as NGC6611. \par
	In summary, optical data in BVI bands have been obtained on 2000 July 29 with WFI/2.2m@ESO, as part of the ESO Imaging Survey (EIS, \citealp{Moma01}), and they have been reduced and calibrated in GPM09. Infrared data in the $JHK$ bands are taken from the Galactic Plane Survey (GPS, \citealp{Luca08}) of the United Kingdom Infrared Deep Sky Survey (UKIDSS), based on observations with the Wide Field Camera (WFCAM, \citealp{Casa07}) on the United Kingdom InfraRed Telescope (UKIRT). Spitzer/IRAC data are taken from the catalog Galactic Legacy Infrared Mid-Plane Survey Extraordinaire (GLIMPSE, \citealp{Benj03}), and they have been analyzed in \citet{Inde07} and GMD09. Both UKIDSS and IRAC data cover the complete WFI field of view (FoV), with the exception of a small region in the northwest for the IRAC data. X-ray data of the center of the Eagle Nebula are taken from the CHANDRA/ACIS-I \citep{Weiss02} observation presented in \citet{Lin07}; in addition, we present here two new CHANDRA/ACIS-I observations pointed toward east and northeast with respect to the cluster (PI: Guarcello). Table \ref{xostable} resumes the three CHANDRA observations. The center of the new observations have been chosen in order to have ColumnV and the embedded cluster (Sect. \ref{m16}) near the centers of the ACIS FoV, where the PSF is narrow and the sensitivity is better. Thereafter, we will call these fields {\it C-field, E-field} and {\it NE-field}, respectively. \par
\begin{figure}[]
	\centering	
	\includegraphics[width=7.5cm]{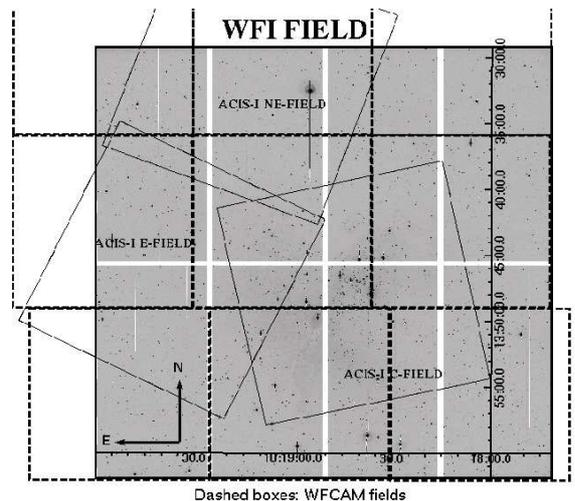}
	\caption{WFI-I image of the Eagle Nebula in $I$ band (the $4\times2$ mosaic). The fields covered by UKIRT/WFCAM (dashed boxes) and ACIS-I (inclined boxes) are also shown. NGC~6611 lies in the center of the ACIS-I $C-field$. North is up, East is on the left.}
	\label{wfifield}
	\end{figure}

Figure \ref{wfifield} shows the $33^{\prime}\times34^{\prime}$ image in $I$ band of the Eagle Nebula, obtained with WFI. The FoVs of WFCAM and ACIS-I are also shown, marked with the dashed and inclined boxes respectively. \par

\subsection{X-ray data}
\label{xraysec}
 
 	\begin{table*}[ht]
	\centering
	\caption {CHANDRA/ACIS-I observations of the Eagle Nebula. The $E$-$field$ has been observed on four different days.}
	\vspace{0.5cm}
	\begin{tabular}{cccccc}
	\hline
	\hline
	Obs. ID& Field name & RA & DEC & Exposure time & Date \\
	       &            &(J2000)& (J2000)& Ksec&   \\	
	\hline
	$978  $    &$C$-$field$ &	$18:18:44.72$&	$-13:47:56.50$&	$78$&	$2001-07-30$\\
	$8931 $    &$NE$-$field$&	$18:19:12.00$&	$-13:33:00.00$&	$80$&	$2008-05-28$\\   
	$8932 $    &$E$-$field$ &	$18:19:36.00$&	$-13:47:24.00$&	$31$&	$2008-06-02$\\   
	$9864 $    &$E$-$field$ &	$18:19:36.00$&	$-13:47:24.00$&	$23$&   $2008-06-07$\\  
	$9865 $    &$E$-$field$ &	$18:19:36.00$&	$-13:47:24.00$&	$17$&	$2008-06-04$\\   
	$9872 $    &$E$-$field$ &	$18:19:36.00$&	$-13:47:24.00$&	$9 $&   $2008-06-09$\\  
	\hline
	\hline
	\multicolumn{5}{l} {} 
	\end{tabular}
	\label{xostable}
	\end{table*}

	The reduction of X-ray data was performed with CIAO 4.0\footnote{http://cxc/harvard.edu/ciao} and the CALDB 4.1.1 calibration files.  In order to perform a consistent study, we analyzed both the new observations in the outer fields and that in the $C$-$field$, which was already analyzed in \citet{Lin07}. For each observation, a {\it Level 2} event file was produced with the {\it ACIS\_PROCESS\_EVENT} CIAO tool from the {\it Level 1} event file. In the Level 2 files we retained only events which produced a charge distribution in the detector that is incompatible with cosmic rays ({\it grades=0,2,3,4,6} and {\it status=0}). The $0.5^{\prime\prime}$ events position randomization added in the Chandra X-ray standard data processing was removed, because it could have affected the sources position determination. \par
The 1 keV exposure map, necessary for point-source detection, was produced with the {\it MERGE\_ALL} task. Point-source detection was performed with {\it PWDetect}\footnote{http://www.astropa.unipa.it/progetti\_ricerca/PWDetect}, a wavelet-based detection algorithm developed at INAF-Osservatorio Astronomico di Palermo \citep{Dami97}. PWDetect found 1158 sources in the $C$-$field$, 363 in the $E$-$field$ and 315 in the $NE$-$field$, with a threshold limit of 4.6$\sigma$, corresponding to 10 expected spurious detections, and considering only photons in the 0.5-8 keV energy band. This $4.6\sigma$ value has been adopted because the expected amount of spurious detections is a negligible percentage with respect to the sources total number, and it is smaller than the number of sources we would lose with higher $\sigma$ values. After a careful visual inspection of the detected sources, we removed 11 double detections of the same source found with wavelets at different spatial scales. Using WEBPIMMS\footnote{http://heasarc.gsfc.nasa.gov/Tools/w3pimms.html} and the relation $L_X$ vs. {\it stellar mass} found by \citet{Prei05}, we evaluated the detection limit in the $C$-$field$ to be just 0.1 solar mass deeper than the corresponding limit in the external ACIS-I fields (from 0.3$\,M_{\odot}$ to $0.2\, M_{\odot}$). A detailed account of the properties of X-ray sources we detected will be the subject of a forthcoming paper. The full list of X-ray sources, compiled with the procedure explained in Appendix \ref{matchapp}, consists of 1755 X-ray sources in the three fields. \par

\subsection{UKIDSS data}
\label{ukidssec}

	Near infrared data in JHK bands were extracted from the UKIDSS/GPS catalog. The GPS survey covers the entire northern and equatorial Galactic plane area of the sky plus the Taurus-Auriga-Perseus molecular complex \citep{Luca08}, and also includes the Eagle Nebula. The observations were made with the UKIRT/WFCAM telescope and consist of short exposures for a total integration time of 80$\,$sec. (J and H bands) and 40$\,$sec. (K band), respectively.  \par 
The WFCAM is equipped with a $2\times2$ array of CCDs, $2048\times2048$ pixels each, with a spatial scale of $0.4^{\prime\prime}$ per pixel, smaller than the 2MASS scale ($1^{\prime\prime}$ per pixel ); the typical FWHM is $0.8^{\prime\prime}$. Infrared observations with such a high spatial resolution are crucial for studies of crowded clusters such as the core of NGC~6611. This high spatial resolution reduces the possibility of confusion in merging the detections in different bands, which is important because spurious matches in this merging process can be mistaken as real stars with non-photospheric colors. \par
	We selected the UKIDSS sources in the WFI FoV with {\it priorsec=0} (sources not duplicated in observations of adjacent fields), {\it merged\_class=-1} (star-shaped sources), {\it JHKerrbit$<$256} (sources with moderate quality warnings\footnote{http://surveys.roe.ac.uk/wsa/index.html}), for a total of 159999 sources down to $J=19^m$. Note that 10\% of sources with a high possibility of being noisy ({\it merged\_class=0}) can be stars and that sources classified as galaxies or probable galaxies ({\it merged\_class=+1,-3}) can be unresolved pairs of point sources. These sources have been excluded from the catalog to produce a more reliable list of cluster members. \par
We matched UKIDSS and 2MASS catalogs of the Eagle Nebula with a $0.3^{\prime\prime}$ matching radius (see Appendix \ref{matchapp}), finding 17396 sources in common between the two catalogs, with few multiple matches. The unmatched sources are 8522 in the 2MASS catalog (mostly bright sources) and 142603 in the UKIDSS catalog (mostly very faint sources). The large number of stars detected only in the UKIDSS reflects the depth of this catalog. \par
\begin{figure}[]
	\centering	
	\includegraphics[width=6cm]{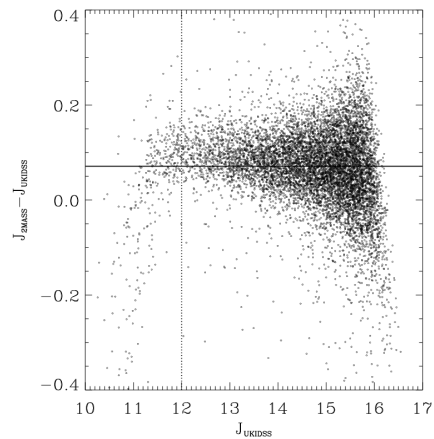}
	\caption{$J_{2MASS}-J_{UKIDSS}$ vs. $J_{UKIDSS}$ for the stars with good photometry. The horizontal line shows the small systematic offset; the dotted line is the adopted $J$ saturation limit.}
	\label{magdiff}
	\end{figure}

	The zero-point calibration of the GPS catalog was performed using 2MASS standard sources \citep{Luca08}. Still, using stars with good photometry in common between the 2MASS and the UKIDSS catalogs, we found small offsets between 2MASS and UKIDSS photometry: $+0.06^m$ (J), $-0.04^m$ (H), $-0.02^m$ (K).  Because magnitudes of UKIDSS/GPS were calibrated in the 2MASS photometric system, we corrected the magnitudes of our UKIDSS sources for these offsets. We also evaluated the saturation limits of our UKIDSS catalog as the magnitude at which the differences between UKIDSS and 2MASS magnitude stars to deviate from the median. The case of the $J$ band is shown in Fig. \ref{magdiff}, where the horizontal line marks the median of the magnitude differences. The limits are $J=12^m$, $H=12.5^m$, $K=11^m$.

\citet{Luca08} claim that the millimagnitude errors quoted in the UKIDSS/GPS catalog are not reliable, and that they should be replaced with a lower limit of $0.02^m$. In this work we follow the suggestion of \citet{Luca08}, so that we do not underestimate the systematic uncertainties.

%%%%%%%%%%%%%%%%%%%%%%%%%%%%%%%%%%%%%%%%%%%%%%%%%%%%%%%%%%%%%%%%%%%%%%%%%%%%%%%%%%%%%%%%%%%%%%%%%%%%%%%%%%%%%%%%%%%%%%%%%%%%%%%%%%%%%%%%%
\section{Multiband catalog}
\label{catsec}

	In the $33^{\prime} \times 34^{\prime}$ region centered on NGC~6611, we selected 28827 optical sources (down to $V=23^m$) 159999 UKIDSS sources (down to $J=19^m$), 41985 IRAC sources (down to $[3.6]=13^m$) and 1755 X-ray sources (down to $F_x=1.5\times 10^{-15}$ $erg \times cm^{-2} \times s^{-1}$). We referred the astrometry of these catalogs to that of 2MASS PSC, and we cross-correlated the sources in these catalogs with the procedure described in Appendix \ref{matchapp}.

	\begin{figure}[]
	\centering	
	\includegraphics[width=9cm]{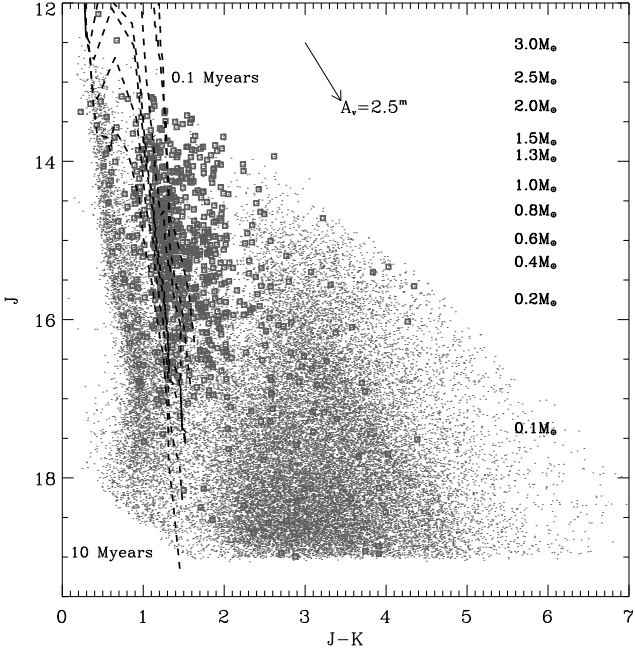}
	\caption{$J$ versus $J-K$ diagram for all stars in WFI FoV (points). The dashed lines are the 0.1, 0.25, 1, 3, 5, 10 Myears isochrones from \citet{Sie00} at the cluster distance and reddening (GPM07); the reddening vector was obtained from the reddening law of \citet{Rie85}. The mass values listed on the right side of the diagrams correspond to stars along the 1 Myear isochrone. The overplotted squares mark the X-ray sources.}
	\label{jjk}
	\end{figure}

Figure \ref{jjk} shows the $J$ versus $J-K$ diagram for all stars in the WFI FoV with both $\sigma_J\leq0.1^m$ and $\sigma_{J-K}\leq0.15^m$. Hereafter, in all the color-color and color-magnitude diagrams shown in this work we will plot only objects which satisfy this color and magnitude selection. The region between the 0.1 and the 3 Myears isochrones (from \citealp{Sie00}) defines the cluster locus, which is well traced by X-ray sources (marked with boxes). The region on the left of the cluster locus is populated by field MS stars. This locus is populated down to the sensitivity limit by sources with both UKIDSS and WFI detection, while those detected also with IRAC (not shown in the figure) populate the field star locus only down to $J\sim15^m$, due to the shallowness of this sample of stars. The region on the right of the cluster locus is that of background giants. Few optical sources fall in this locus, which is dominated by only UKIDSS sources down to the sensitivity limit. \par

%%%%%%%%%%%%%%%%%%%%%%%%%%%%%%%%%%%%%%%%%%%%%%%%%%%%%%%%%%%%%%%%%%%%%%%%%%%%%%%%%%%%%%%%%%%%%%%%%%%%%%%%%%%%%%%%%%%%%%%%%%%%%%%%%%%%%%%%%

\section{Candidate members with circumstellar disk}
\label{disksec}

	\subsection{YSOs from IRAC colors}
	\label{glimpsesec}

	\begin{figure*}[]
	\centering	
	\includegraphics[width=13cm]{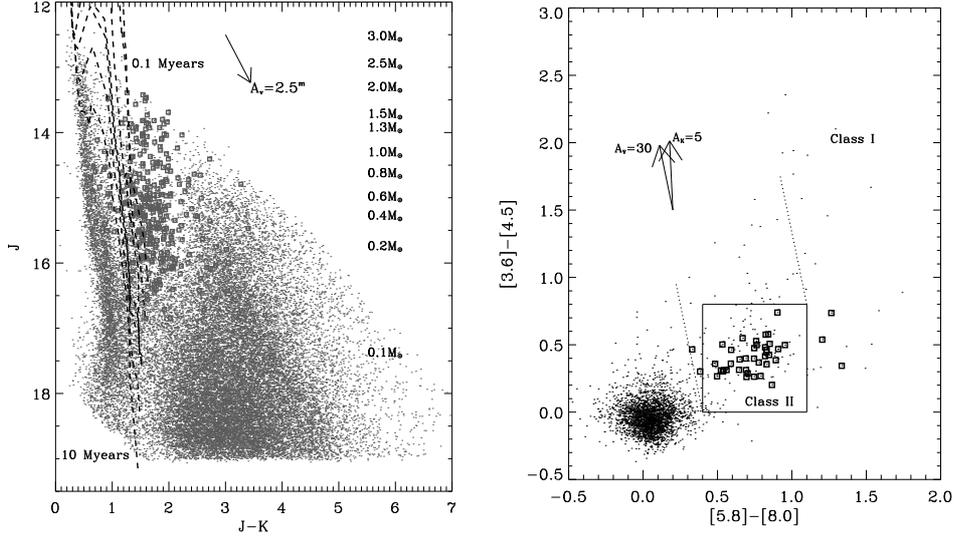}
	\caption{Color-magnitude diagrams for the stars (gray points) in the WFI FoV ($J$ vs $J-K$ in the left and the IRAC color-color diagram in the right panel). Squares mark the subsample of stars with excesses in IRAC bands selected with $Q_{VIJ[sp]}$ indices with good UKIDSS photometry (left panel) and good IRAC photometry (right panel). In the left panel the ZAMS and isochrones are from \citet{Sie00}. The mass values listed on the right side of the diagrams correspond to stars along the 1 Myear isochrone. In the right panel the box delimits the approximate locus of ClassII YSOs \citep{Alle04}. The dashed lines in the right panel roughly separate reddened photospheres (left), reddened ClassII YSOs (center), and ClassI YSOs. The reddening vectors in the right panel with A$_V$=30$^m$ and A$_K$=5$^m$ are from \citet{Mege04} and \citet{Fla07}, respectively.}
	\label{dagqirfig}
	\end{figure*}

Candidate young stars with infrared excesses have been selected using two criteria: IRAC color-color diagram and suitable reddening-free color indices ($Q$ indices). \par
	The selection of disk-bearing YSOs with the IRAC color-color diagram has been already presented in GMD09, who found 143 ClassII stars, 13 ClassI, and 22 objects not clearly classified (ClassI or highly embedded ClassII stars). Figure 3 of GMD09 shows the IRAC color-color diagram and the selection of YSOs associated with M16.  \par
Stars with excesses in the IRAC bands were also selected using the $Q$ color indices, which are independent from interstellar extinction. $Q$ indices have been described in detail in \citet{Dami06}, GPM07 and GMD09. In particular, the $Q$ indices used to select stars with excesses in IRAC bands, $Q_{VIJ[sp]}$, are defined as

	\begin{equation}
	Q_{VIJ[sp]} = \left( V-I \right) - \left( J-[sp] \right) \times E_{V-I}/E_{J-[sp]},
	\label{Qdef}
	\end{equation}
	\par

where $[sp]$ is the magnitude in an IRAC band and $V-I$ is representative of photospheric emission. In Eq. \ref{Qdef} the reddening $E_{V-I}$ is obtained from \citet{Muna96}, $E_{J-[sp]}$ was inferred from the reddening law of \citet{Mat90}. In the $Q_{VIJ[sp]}$ vs. $(J-[sp])$ diagram (see GMD09), stars with excesses in the $[sp]$ band have smaller indices than those of normal stars, while the extinction vector is horizontal. This allows the separation of stars with excesses from extincted sources. In the $Q_{VIJ[sp]}$ vs. $(J-[sp])$ diagrams, stars with excesses are defined by us as those with the $Q$ index smaller (below $3\times \sigma_Q$) than the limit for the photospheric emission. We select 343 candidate stars with disks considering $Q_{VIJ[sp]}$ indices, which were mostly already found with the 2MASS photometry (the selection of stars with disks with UKIDSS and 2MASS will be compared in Sect. \ref{ukvs2msec}). The selection of stars with excesses in IRAC bands that we performed here with $Q$ indices is only slightly different from that in GMD09, because we use here the UKIDSS $J$ band, instead of the 2MASS $J$. Both selections share the same lower limit in $J$ ($\sim16^m$), due to the completeness limit of IRAC data, but have a different upper limit ($J\sim 13.5^m$ with $J_{UKIDSS}$, $J\sim 10^m$ with $J_{2MASS}$). In GMD09 the selection of YSOs with the IRAC color-color diagrams is compared in detail with that of \citet{Inde07}.

The nature of the $Q_{IRAC}-excess$ stars is shown in the diagrams in Fig. \ref{dagqirfig}. In the $J$ vs $J-K$ diagram (left panel), the $Q_{IRAC}-excess$ stars lie in the cluster locus with $J-K$ redder than the colors corresponding to the 0.1 Myear isochrone, likely due to a $K$ excess. These stars lie in the cluster locus also in the optical color-magnitude diagrams. Because $Q_{VIJ[sp]}$ requires the combination of WFI, UKIDSS and IRAC photometry, this selection cannot be extended to the faint UKIDSS sources. The IRAC color-color diagram (right panel) shows that all these stars are ClassII objects, mostly with low reddening.

	\subsection{YSOs from UKIDSS colors}
	\label{uksec}
Stars with excesses in the UKIDSS bands have been selected with the $Q$ indices defined in GPM09. Four indices compare $V-I$ with NIR colors:

	\begin{equation}
	Q_{UKIDSS} = \left( V-I \right) - \left( A-B \right) \times E_{V-I}/E_{A-B},
	\label{Qukdef}
	\end{equation}
	\par

where $A-B$ assumes the values $I-J$, $J-H$, $J-K$ or $H-K$ and the reddening $E_{A-B}$ is calculated according to \citet{Rie85} and \citet{Muna96}. Another index compares $J-H$ with $H-K$:

	\begin{equation}
	Q_{JHHK} = \left( J-H \right) - \left( H-K \right) \times E_{J-H}/E_{H-K},
	\label{Qjhhkdef}
	\end{equation}
	\par

\begin{figure}[!h]
	\centering	
	\includegraphics[width=8cm]{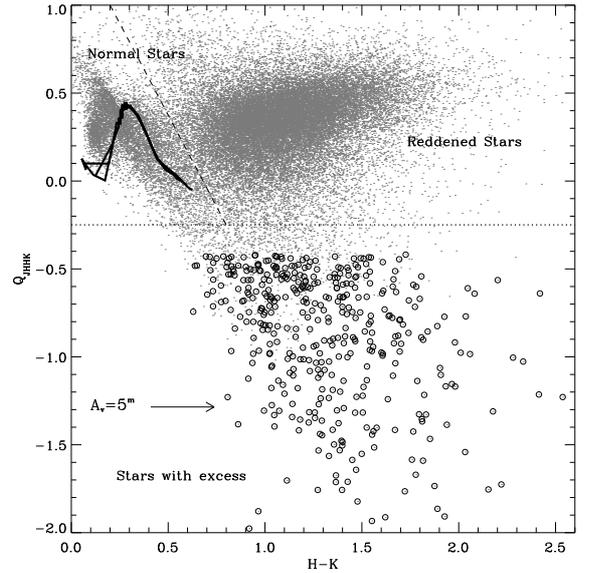}
	\caption{$Q_{JHHK}$ vs. $H-K$ diagram for the stars in the WFI FoV (gray points). Dashed and dotted lines separate the loci of normal stars, reddened stars, and stars with excess (circles). The solid lines are the isochrones as in Fig. \ref{jjk}.}
	\label{Qjhhkfig}
	\end{figure}

Because of detection in the optical bands is not required, this index allows us to select more absorbed stars than the $Q_{VIJ[sp]}$ and the other $Q_{UKIDSS}$ indices. Figure \ref{Qjhhkfig} shows the $Q_{JHHK}$ vs. $H-K$ diagram for stars in the WFI FoV (gray points), which has been used to select stars with excesses in $K$ band (marked with circles). These stars are those with a more negative index than the corresponding photospheric limit by more than 3$\sigma_Q$. The other $Q_{VIJ[sp]}$ and $Q_{UKIDSS}$ diagrams are analogous. 

 	\begin{table}[ht]
	\centering
	\caption {Number of stars with excesses in $JHK$ bands selected with the $Q$ indices defined with UKIDSS (second column) and 2MASS (third column) photometric data.}
	\vspace{0.5cm}
	\begin{tabular}{ccc}
	\hline
	\hline
	$Q$ index & UKIDSS detection & 2MASS detection \\
	\hline
	$Q_{JHHK}$&	$448$&	$22 $\\
	$Q_{VIIJ}$&	$39 $&	$34 $\\   
	$Q_{VIJH}$&	$375$&	$91 $\\   
	$Q_{VIJK}$&	$751$&	$159$\\  
	$Q_{VIHK}$&	$609$&	$136$\\   
	\hline
	\hline
	\multicolumn{3}{l} {} 
	\end{tabular}
	\label{quk2mtable}
	\end{table}

Table \ref{quk2mtable} compares the number of candidate stars with excesses in $JHK$ bands selected with UKIDSS (this work) and 2MASS photometry (GPM09); the number of selected stars is 1264\footnote{The number of stars selected with $Q_{UKIDSS}$ is weakly depending on the adopted saturation limits (see Sec. \ref{ukidssec})): increasing the limits by one magnitude we selected two more stars, decreasing them by one magnitude we selected 29 stars less.} and 345, respectively. This large difference cannot be explained only by the depth of the UKIDSS catalog.

\begin{figure*}[]
	\centering	
	\includegraphics[width=13cm]{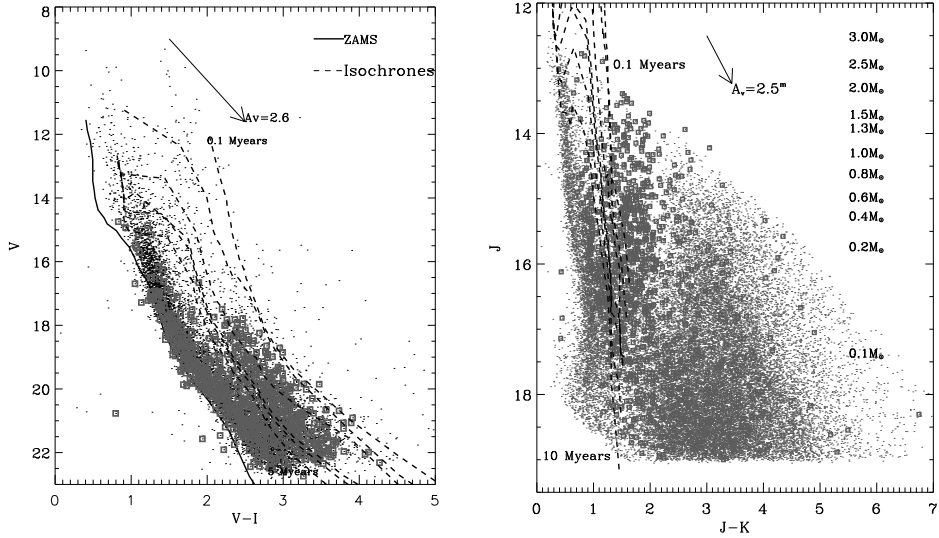}
	\caption{$V$ vs. $V-I$ and $J$ vs. $J-K$ diagrams for all stars in the WFI FoV (gray points) and all stars with excesses in UKIDSS bands selected with $Q_{UKIDSS}$ indices (squares). Isochrones and masses as in Fig. \ref{jjk}}
	\label{diagqukfig}
	\end{figure*}

Figure \ref{diagqukfig} shows the $V$ vs. $V-I$ and $J$ vs. $J-K$ diagrams for all the stars in the WFI FoV (gray points) and the stars with excesses selected with the $Q_{UKIDSS}$ indices. In the right panel a significant number of these stars are faint and highly reddened. In the left panel it is evident that a great number of $Q_{UKIDSS}-excess$ stars lie in the locus of field stars instead of in the cluster locus. It is possible that a physical mechanism, such as gas accretion from the disk to the central star or scattering of stellar light into the line of sight by disk's grains, could produce these blue {\it optical} colors in young disk-bearing stars, which fall in the field locus in the various optical color-magnitude diagrams (\citealt{Har90}; \citealt{Hille97}; \citealt{Io10}). These effects, however, affect only the optical colors, while most of these stars fall in the field star locus also in diagrams defined with NIR photometry. This suggests that some stars have been wrongly selected with $Q_{UKIDSS}$ indices. We decided, then, to confirm as disk-bearing cluster members the $Q_{UKIDSS}-excess$ stars which share at least one of the following features:

\begin{enumerate}
\item they are compatible with PMS stars in the IRAC color-magnitude diagrams (defined by two or three IRAC bands), in the $V$ vs. $V-I$, $I$ vs. $I-J$ or $J$ vs. $J-K$ diagrams (424 stars);
\item they are also $Q_{IRAC}-excess$ stars (102 stars);
\item they fall in the ClassII locus in the IRAC color-color diagram (49 stars).
\end{enumerate}

	Adopting these criteria, we selected 575 candidate disk-bearing stars, down to 0.2 $M_{\odot}$. \par
We discarded 399 $Q_{UKIDSS}-excess$ stars from the list of candidate members because

\begin{enumerate}
\item they lie in the field-stars loci both in the optical and in the infrared color-magnitude diagrams (389 stars);
\item or they have normal colors in the $Q_{VIJ[sp]}$ vs. $(J-[sp])$ diagrams (10 stars);
\end{enumerate}

We have no firm explanation for the 399 objects for which the $Q$ selection seems to fail. One possibility is that these stars are false matches between the optical and infrared catalogs. However, the expected amount of spurious identifications, evaluated as in \citet{Dami06}, is 107, 3.7 times smaller than the amount of false $Q$ selections. More likely, these selections are induced because the adopted reddening laws is not reliable, modifying the ratio of color excesses used in Eqs.  \ref{Qukdef} and \ref{Qjhhkdef}. This should happen both because there can be problems in the calibration of UKIDSS/GPS data on the 2MASS photometric system, especially for very reddened objects \citep{Luca08}, and because the adopted extinction laws are defined for the usual interstellar $R_V=3.1$, while in the direction of NGC~6611 the reddening law is anomalous (the value $R_V=3.1$ is used for the ratio between color excesses in $Q$ definitions, even if a slightly higher value, 3.27, has been found in GPM07). Different values of $R_V$ change both the slope of the reddening vectors in the color-magnitude diagrams and the relations between the reddening in various colors. The reddening laws we used do not give the exact relations between the reddening at different bands if $R_V$ differs from the usual 3.1.  \par
	The remaining 290 stars are retained in the final catalog as ``{\it probable members}'', because we could not assess their nature, with a flag indicating their ``uncertain members status''. Most of them are faint $Q_{JHHK}-excess$ sources without any other counterpart. We retained these stars because most of the new detections of low-mass disk-bearing stars provided by UKIDSS should belong to this group.\par

	\subsection{Comparison between UKIDSS-based and 2MASS-based disk diagnostics}
	\label{ukvs2msec}

Comparing the list of candidate stars with excesses in $JHK$ bands selected with $Q_{2MASS}$ in GPM07 and the list of stars with reliable excesses selected here with $Q_{UKIDSS}$, we found

\begin{itemize}
\item 139 sources selected only with $Q_{2MASS}$;
\item 391 sources selected only with $Q_{UKIDSS}$, among which 307 are listed in the 2MASS Point Sources Catalog;
\item 184 sources in both lists.
\end{itemize}

\begin{figure}[!h]
	\centering	
	\includegraphics[width=9cm]{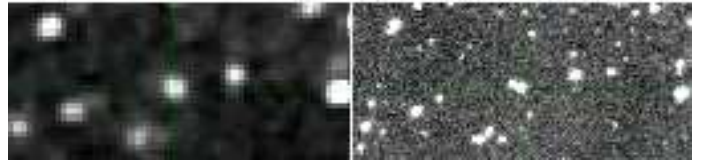}
	\caption{UKIDSS (right panel) and 2MASS (left panel) images of a disk-less star taken from our multiband catalog corresponding to a 2MASS source with  $K$ band excess according to 2MASS data.}
	\label{18534fig}
	\end{figure}

Among the stars of the first group, 52 have good UKIDSS photometry but do not have excesses in $Q_{UKIDSS}$ diagrams. We compared the UKIDSS and 2MASS images for each of these stars and verified that the excesses found in 2MASS bands for all the stars of this group are due not to a disk, but to one or more close stars resolved in UKIDSS images but not in 2MASS. As an example, Fig. \ref{18534fig} shows one star which belongs to such a group. These 52 stars cannot anymore be considered disk-bearing stars and therefore cluster members.

	\subsection{Contamination of the list of candidate members with disk}
	\label{contsec}

Because not all selected candidate members with disk are X-rays sources, a significant contamination by both foreground and background sources can be expected. Foreground contaminating sources are expected to dominate the sample of 399 stars discarded from the list of candidate members with disk (see Sect. \ref{uksec}), by reason of their optical colors. We retained in our final list of candidate stars with disk only $\sim\,90$ stars with infrared excesses and optical colors similar to those of foreground stars. In \citealt{Io10} their nature is discussed in detail, and we conclude in that paper that at least 2/3 of these stars are bona-fide YSOs associated to M16. \par 
	The identification of candidate background contaminating sources is a hard task, because they can be confused with largely embedded YSOs associated to M16. However, we do not expect a significant amount of background sources among the candidate members with disks detected in the optical bands, because the cloud absorbs the background optical radiation efficiently. This is evident in the $V$ vs. $V-I$ diagram, where stars in the Main Sequence locus are observed up to the distance of M16. In order to estimate the amount of visual extinction of background sources due to M16, we used the UKIDSS sources without optical counterparts, which are dominated by background giant stars, because their intrinsic colors can be accurately determined. We divided the WFI fields in 400 cells, in which fall between 37 and 652 sources, accounting for a good statistic. We evaluate the median $H-K$ of the sample of sources falling in each cell and then the median $E_{H-K}$ for each cell, taking advantage of the fact that the intrinsic $H-K$ colors for giant stars range from 0.07 to 0.31, with a median of 0.2 \citep{Besse88}. Because the sample of UKIDSS sources without optical emission is populated by giant stars at large distance (up to $\sim \,10\,Kpc$, as suggested by \citealp{Luca08}), the values of $A_V$ we obtain with this procedure is strongly affected by the contribution of the material in the Galactic plane. However, the lowest values we obtain ($8^m<A_V<10^m$) should to be related to the background giant stars which are closest to M16 (and then mostly due to the cloud). With this estimate of visual extinction for the background stars immediately behind the nebula, we can understand whether a background giant star can be observed in our optical observation and can contaminate, for some reason, our list of candidate members with disk. To this aim, we consider the $V-K$ color of a G0 giant star (the brightest type of giant star listed in \citealt{Besse88}), which at a distance of $1800\,pc$ and with an extinction ranging from $8^m$ to $10^m$ is between $8.35^m$ and $10.63^m$. Only 17 candidate stars with disk have the $V-K$ color in this range, so we conclude that the contamination by background stars of our list of candidate members with disk with optical detection is negligible. \par
Background giant stars and galaxies observed through the Galactic plane can contaminate the sample of candidate stars with disk which are not detected in optical, because of the depth of the UKIDSS catalog. However, as explained in Sect. \ref{uksec}, all these stars, except for 21, were classified as ``probable cluster members'' and not used in the study of disk frequency and star-formation chronology. In this case, as well, the contamination by background sources is negligible. This is also valid for the YSOs selected with the IRAC color-color diagram, as explained in GMD09. \par
	We then can safely assume that our sample of candidate stars with disks is not strongly affected by contamination from either foreground or background sources.
%%%%%%%%%%%%%%%%%%%%%%%%%%%%%%%%%%%%%%%%%%%%%%%%%%%%%%%%%%%%%%%%%%%%%%%%%%%%%%%%%%%%%%%%%%%%%%%%%%%%%%%%%%%%%%%%%%%%%%%%%%%%%%%%%%%%%%%%%%5

\section{Final list of cluster members}
\label{listsec}

	We decided to compile the list of stars with excesses in $JHK$ including all stars selected with $Q_{UKIDSS}$ with reliable excesses (see Sect.\ref{uksec}), plus the stars with only 2MASS excesses with saturated UKIDSS photometric quality flags and those with bad UKIDSS photometry. Considering also the stars with excesses in IRAC bands, candidate stars with disks can be classified as follows

\begin{figure*}[]
	\centering	
	\includegraphics[width=13cm]{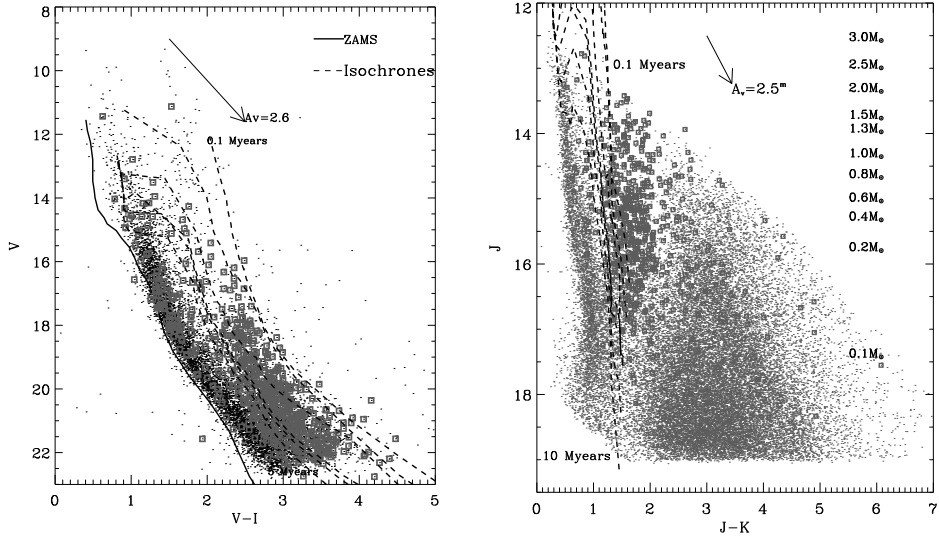}
	\caption{$V$ vs. $V-I$ and $J$ vs. $J-K$ diagrams for all stars in the WFI FoV (gray points) and all candidate members with disk (squares).}
	\label{diagqtenfig}
	\end{figure*}

\begin{itemize}
\item 627 stars selected with $Q$ indices, among which
	\begin{itemize}
	\item 269 stars selected with both $Q_{IRAC}$ and $Q_{UKIDSS}$;
	\item 74 stars selected only with $Q_{IRAC}$;
	\item 284 stars selected only with $Q_{UKIDSS}$.
	\end{itemize}
\item 178 stars selected with the IRAC color-color diagram, among which
	\begin{itemize}
	\item 52 stars selected both with the $Q$ indices and the IRAC diagram;
	\item 126 stars selected only with the IRAC color-color diagram.  
	\end{itemize}
\item 81 stars added from those selected with $Q_{2MASS}$ indices, which cannot be selected by $Q_{UKIDSS}$ since
	\begin{itemize}
	\item they are brighter than the saturation limits (5 stars);
	\item bad UKIDSS photometric quality flags (76 stars).
	\end{itemize}
\end{itemize}

	for a total of 834 (627+126+81) candidate cluster members with disk (Fig. \ref{diagqtenfig} shows the $V$ vs. $V-I$ and $J$ vs. $J-K$ diagrams with the selected candidate stars with disk). These stars with infrared excesses can be divided into three groups.

\begin{itemize}

\item Candidate ClassII objects, from which we observe both the emission from the photosphere and from the inner region of the disk. Because none among these stars fall in the {\it normal stars} locus in some of the $Q$ diagrams\footnote{they fall in the loci of stars with excesses or those where excess and reddening cannot be discerned.}, the excesses in all the NIR bands should come from the same region of the disk, i.e. the inner rim at the dusts sublimation radius. 
\item Candidate ClassII objects selected only with $Q_{VIJ[8.0]}$ (4 stars), which likely have dissipated their inner disks and their Spectral Energy Distributions (SEDs) have a broad silicate band in emission, responsible for the excess in [8.0].
\item Candidate YSOs selected only with the IRAC color-color diagram, given the lack of detection in optical or $J$ bands. These are likely embedded ClassII and ClassI YSOs. $Q$ indices are useless to select these stars, while the IRAC color-color diagram is an efficient tool to this end.
\end{itemize}
	
The final list of cluster members includes the selected candidate stars with disk (834 stars) and the candidate young stars associated with M16 without a disk. These stars are selected from the X-ray sources which meet the following requirements:

\begin{itemize}
\item they have at least one stellar counterpart;
\item the positions of the stellar counterparts in the $V$ vs. $V-I$ and $J$ vs. $J-K$ diagrams have to be compatible with the cluster locus.
\end{itemize}

The last requirement discards all the foreground contaminants sources (72), whose position in the color-magnitude diagrams cannot lie in the cluster locus. The first requirement allows us to discard almost all background contaminating sources, whose optical or infrared emission cannot be observed because of the absorption by the interstellar material along the galactic plane. We estimated the maximum number of expected extragalactic sources observed in our ACIS observations taking advantage of the $logN\,vs.\,logS$ diagrams of  \citet{Pucce06}, which is a work based on XMM-Newton observations of the extragalactic sources lying in the ELAIS fields. In order to compare our data with their $logN\,vs.\,logS$ diagrams, we converted\footnote{To this aim we used WEBPIMMS} our limit counts rate ($6.25 \times 10^{-5}$ cnts/sec) into flux in the 0.5-10 keV energy band with a power-law model with energy index $\alpha$=0.8 in order to be consistent with \citet{Pucce06}. We set the interstellar extinction $nH$ equal to $5.4 \times10^{22}\,cm^{-2}$ from the highest $A_V$ ($30^m$) we can estimate from the sample of UKIDSS sources without optical detection, which is dominated by background giants stars observed up to large distances (see Sect. \ref{contsec}). With this method, we estimate $\sim193.5$  extragalactic sources to be detected in our ACIS-I observations. But this is clearly an upper value to the real number, because we expect further intervening extinction (i.e. a larger nH) even behind the M16 nebula, which reduces the number of detectable extragalactic sources in our ACIS-I images, and also for the decrease of limiting sensitivity at large off-axis angle in the ACIS-I detector, which has not been properly taken into account in the above estimate. Since there are 504 X-ray sources without stellar counterparts in our catalog, it is possible to assume that about 300 among them are very embedded YSOs associated with M16 instead of contaminating background objects. \par

	The total of candidate disk-less pre-main sequence stars associated to the Eagle Nebula is then equal to 1103 sources. This is not a complete census of the X-rays-emitting sources of M16, because we lost all quiescent sources with a luminosity $L_X$ below our threshold (which is $log(L_X)=29.8\,erg/s$ at $1750\,pc$) during the observation. To estimate the percentage of cluster members at various masses that we do not observe, we used the existing most complete census of the X-ray population of a 1 Myear cluster: the Chandra Ultradeep Orion Project ($COUP$ \citealp{Get05}). \citet{Fei05} calculated the unabsorbed $L_X$ distribution in the whole ACIS-I band sampling the Orion sources in five different mass bins. With their results and comparing the distributions with our detection limit, we evaluated that we observed 42\% of stars with a mass smaller than 0.3 $M_{\odot}$ (lower than the limit of our optical and IRAC data), 83\% of stars with masses between 0.3 and 1 $M_{\odot}$ , 92\% between 1 and 3 $M_{\odot}$ , 100\% between 3 and 10 $M_{\odot}$  and 75\% between the stars more massive than 10 $M_{\odot}$. We can estimate then that we observed about the 86\% of clusters members more massive that 0.3 $M_{\odot}$, which is the limit of our optical and IRAC data (and which is the limit of the members selected to study the spatial variation of disks frequency and the chronology of star formation in the following sections).

%%%%%%%%%%%%%%%%%%%%%%%%%%%%%%%%%%%%%%%%%%%%%%%%%%%%%%%%%%%%%%%%%%%%%%%%%%%%%%%%%%%%%%%%%%%%%%%%%%%%%%%%%%%%%%%%%%%%%%%%%%%%%%%%%%%%%%%%%

\section{Spatial variation of disk frequency}
\label{diskfresec}

\begin{figure*}[]
	\centering	
	\includegraphics[width=6cm]{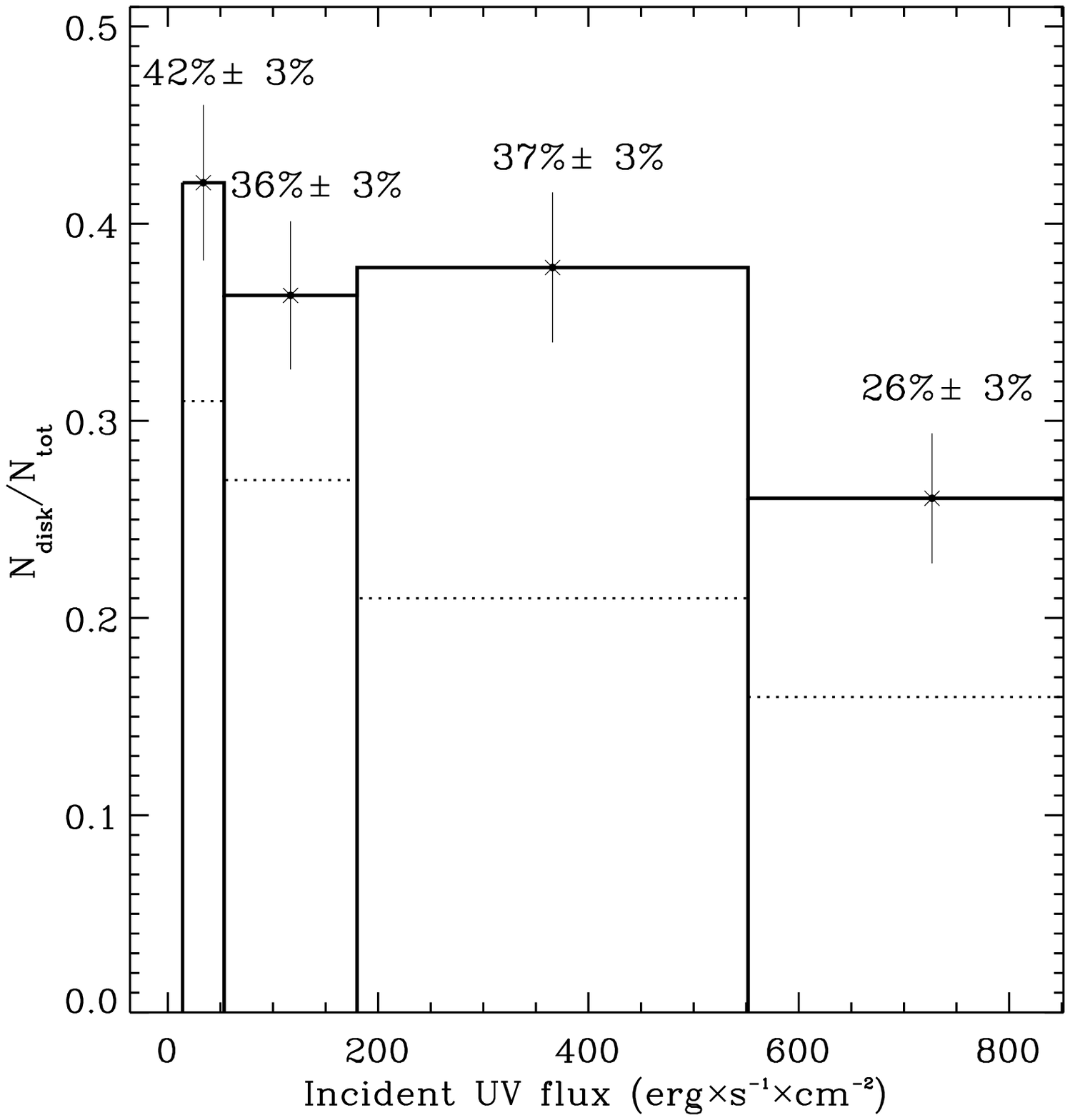}
	\includegraphics[width=6cm]{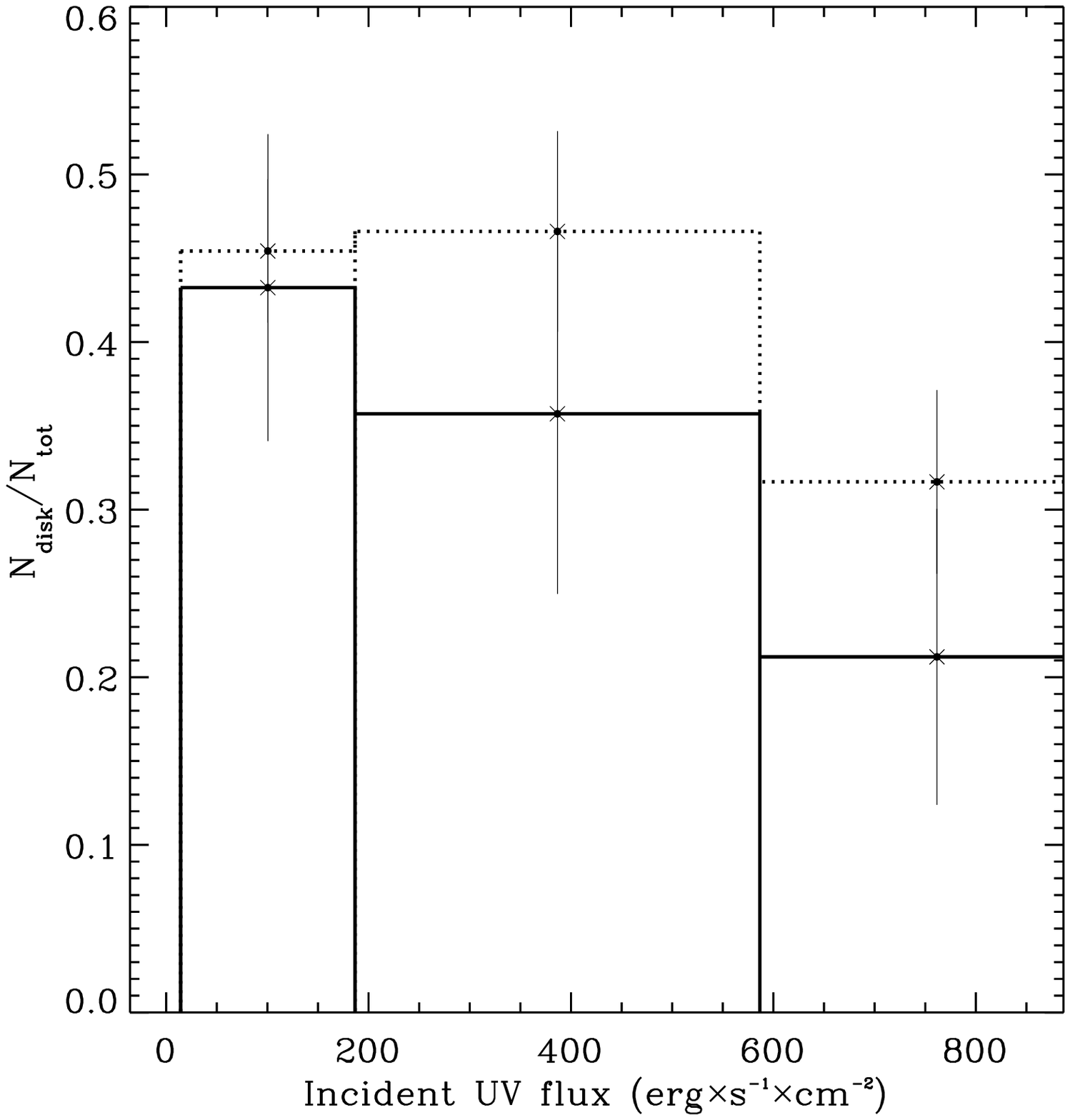}
	\caption{Left panel: histogram of the disk frequency in NGC~6611 (the $C$-$field$) vs. the flux emitted by massive stars and incident on cluster members, compared with that found in GMD09 (dotted histogram). Right panel: histograms analogous to that in the left panel, but with the cluster members in two different ranges of stellar masses: $>1M_{\odot}$ (solid line) and $0.2 M_{\odot} < M_{stars} < 1 M_{\odot}$ (dotted line).}
	\label{istofig}
	\end{figure*}

	Disk frequency is expected to vary across the Eagle Nebula given the presence of recent star-formation sites and the not uniform spatial distribution of massive stars, which are mostly concentrated in the central cavity corresponding to NGC~6611. The spatial variation of disk frequency in the central region (the {\it C-field}) has been studied in GPM09 and GMD09, where it was found that the disk frequency decreases close to massive stars. These studies pertained only to the $C$-$field$, which contains almost all the massive stars of NGC~6611. In this field, we selected here 896 candidate PMS members without disk and 510 candidate cluster members with disk. The new list of cluster members compiled in this work with the UKIDSS data confirms the decrease of the disk frequency close to massive stars, as shown in the histogram (left panel of Fig. \ref{istofig}) of the {\it disk frequency} vs. {\it the flux emitted by OB stars and incident on cluster members}. The incident fluxes (see GPM07 for detail on flux evaluation) were calculated by summing the emission from all massive stars in the cluster and incident on each cluster members brighter than $J=17^m$ (i.e. with masses larger than $0.1-0.2\,M_{\odot}$, in order to explore the same mass ranges of disk-less and disk-bearing candidate members). The dotted lines in the left panel of Fig. \ref{istofig} show the disk frequencies in each UV flux bin obtained in GMD09. The disk frequencies found in this work are higher by $\sim$10\% than those found in GMD09. This reflects the larger number of candidate disk-bearing stars we selected with UKIDSS. The average disk frequency in the whole $C$-$field$ is now equal to $36\% \pm 1\%$ ($33\% \pm 2\%$ taking into account the completeness of our X-ray survey for stars more massive than 0.3$\,M_{\odot}$, see Sect. \ref{listsec}).

The visual extinction map, discussed in Appendix \ref{extapp}, allowed us to obtain the dereddened $V_0$ vs $\left(V-I \right)_0$ and $J_0$ vs $\left(J-K \right)_0$ diagrams. Thanks to these diagrams, we estimated the masses of the cluster members from the tracks of \citet{Sie00} and we studied the disk frequency at different mass ranges. Table \ref{diskfreqtable} shows the average disk frequency in the $C$-$field$ (which contains NGC~6611) for two different mass ranges. There is no large dependence of the disk frequency on the stellar mass, even if a slightly ($1.5 \times \sigma_{freq}$) higher frequency is found for masses $\leq 1 M_{\odot}$. This difference becomes even less significant taking into account the percentage of X-ray sources below our $L_X$ threshold (Sect. \ref{listsec}). The right panel in Fig. \ref{istofig} shows the disk frequency for two different ranges of central star masses (larger and smaller than $1 M_{\odot}$, respectively) and for different values of incident fluxes emitted by massive stars, taking into account only the candidate members that fall inside the $C$-$field$. The difference between the disk frequencies for high- and low-mass stars does not strongly depend on the intensity of the incident flux. Note that the disk frequencies in Table \ref{diskfreqtable} differ slightly from the average frequency in NGC~6611 ($36\% \pm 1\%$), because it was impossible to evaluate the masses of all candidate cluster members that fall in the $C$-$field$.

	\begin{table}[h]
	\centering
	\caption {Disk frequency in the $C$-$field$ for different mass ranges}
	\vspace{0.5cm}
	\begin{tabular}{cccc}
	\hline
	\hline
	Mass range & Disk-less & Disk-bearing & Disk frequency \\
	\hline
	$> 1 M_{\odot}$&	$119$&	$56 $&	$32\% \pm 4\%$\\
	0.3-1 $M_{\odot}$&	$394$&	$300$&	$43\% \pm 3\%$\\
	\hline
	\hline
	\multicolumn{4}{l} {} 
	\end{tabular}
	\label{diskfreqtable}
	\end{table}

\begin{figure}[]
	\centering	
	\includegraphics[width=9cm]{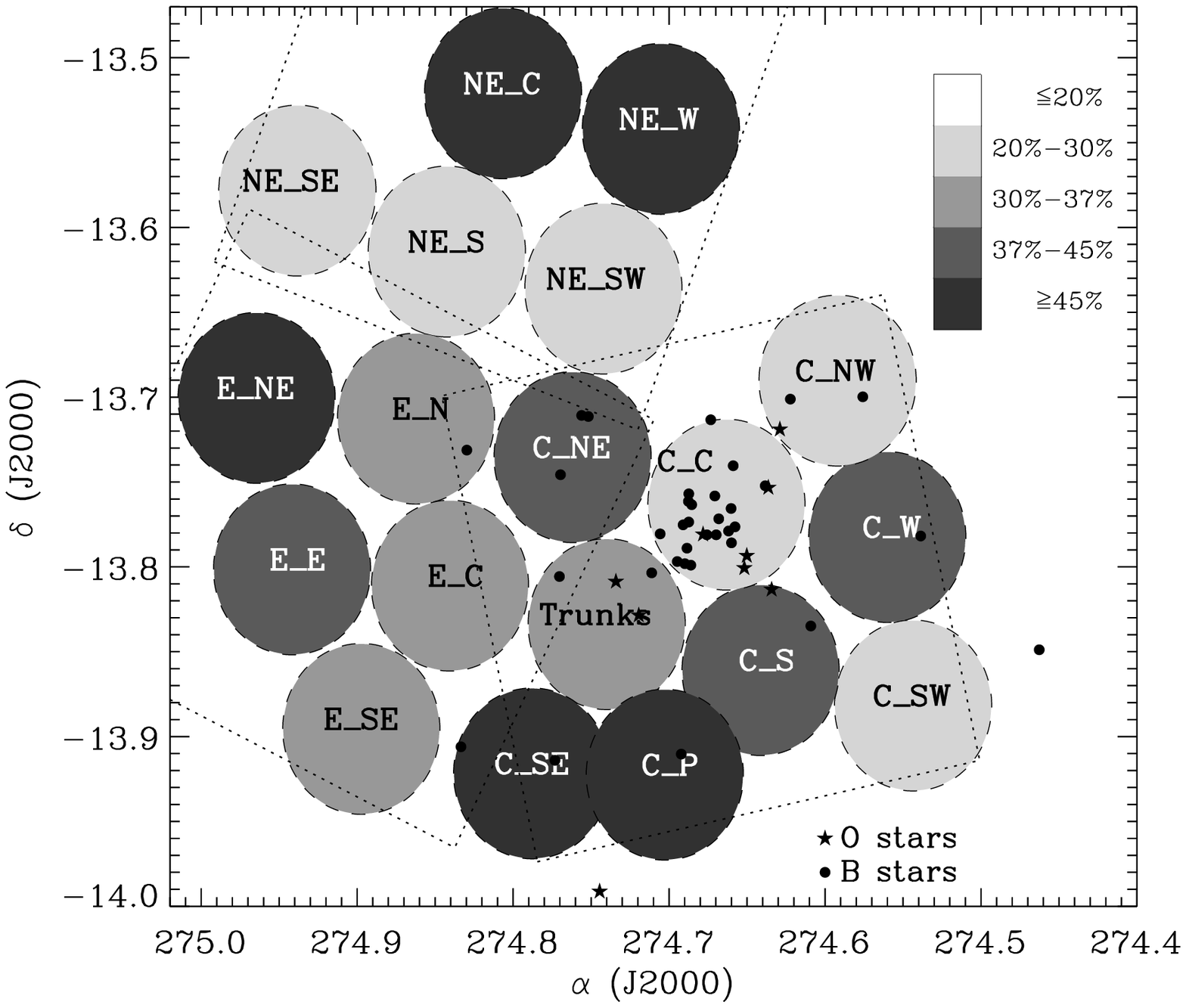}
	\caption{Gray-scale map of the disk frequency in M16. Dots and stars mark the positions of OB (earlier than B5) stars. The dotted boxes are the ACIS-I fields. The circles mark the regions we defined.}
	\label{diskfreqmap}
	\end{figure}

	In order to compare with a good statistic the spatial variation of disk frequency in the outer regions of M16 with the morphology of the nebula, the average age of cluster members, and the positions of OB stars, we defined 20 regions across the portion of the Eagle Nebula observed with ACIS-I. Figure \ref{diskfreqmap} shows these regions and their names, the disk frequencies (on a gray scale) and their content of OB stars.  Each region is labeled with the name of the ACIS-I field in which it is contained (the first letters, one or two), and the orientation with respect to the center of this ACIS-I field (for instance, the region NE\_S is the region south of the $NE$-$field$). Figure \ref{diskfreqmap} shows that the disk frequency is higher in specific outer regions of the nebula with respect to NGC~6611 (mostly contained in the $C\_C\,field$). This could be due either to the presence of a younger population of members or to a lack of OB stars in the outer regions. The disk frequency in all regions varies from $52\% \pm 13\%$ (C\_SE region) to $17\% \pm 8\%$ (E\_SE region). \par

Finally, as explained in Sect. \ref{uksec}, we used $R_V=3.1$ in the definitions of $Q$ indices. However, it is well known that the reddening law in the direction of NGC~6611 is anomalous, even if GPM07 revised the estimate of $R_V$, finding a value (3.27) only slightly different from the canonical one. Performing the selection of disks with $R_V=3.7$, the number of selected candidate members with disks decreases to 8\%, and the average disk fraction to 3\%. The spatial variation of the disk frequency and its relation with stellar masses remains essentially unchanged.

%%%%%%%%%%%%%%%%%%%%%%%%%%%%%%%%%%%%%%%%%%%%%%%%%%%%%%%%%%%%%%%%%%%%%%%%%%%%%%%%%%%%%%%%%%%%%%%%%%%%%%%%%%%%%%%%%%%%%%%%%%%%%%%%%%%%%%%%%

\section{Chronology of star formation}
\label{agesec}

	In order to shed some light on the star-formation history in the Eagle Nebula and to understand if the disk frequency correlates with the age of the stellar population, we evaluated the age of each disk-less member from the isochrones of \citet{Sie00}, with the $V_0$ vs $\left(V-I \right)_0$ and $J_0$ vs $\left(J-K \right)_0$ diagrams. We excluded disk-bearing members from this computation, because their colors were not purely photospheric. A total of 333 disk-less members can be plotted in both diagrams, obtaining two different age estimates for these stars. On average, optical and infrared age estimates for each star agree, with a median difference of $\sim 0.02$ Myears but with a large spread. The age of these stars was then set equal to the average value of the two estimates.  \par

	\begin{figure}[!h]
	\centering	
	\includegraphics[width=9cm]{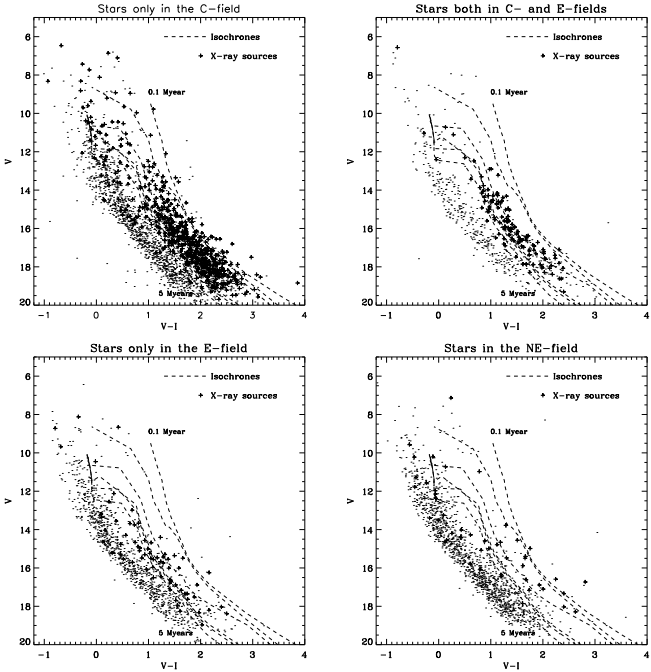}
	\caption{Intrinsic $V_0$ vs. $(V-I)_0$ diagrams for the optical sources (points) and the optical+X-ray sources (crosses) that fall in four different sky regions: only in the $C$-$field$ (upper left panel), in the common region of $C$-$field$ and $E$-$field$ (upper right panel), only in the $E$-$field$ (lower left panel) and falling in the $NE$-$field$ (lower right panel). The isochrones are from \citet{Sie00}.}
	\label{vvidiag}
	\end{figure}

Figure \ref{vvidiag} shows the dereddened $V_0$ vs. $(V-I)_0$ diagrams for the stars falling only in the $C$-$field$ (upper left panel); in the sky region common to $C$-$field$ and $E$-$field$ (upper right panel); only in the $E$-$field$ (lower left panel); and, finally, for the sources falling in the $NE$-$field$. The sequence upper-left, upper-right, lower-left panels go through the nebula from NGC~6611 eastward. In all these diagrams, crosses mark the X-ray sources. The disk-less members, traced by X-ray active stars, in the upper right panel (region in common between the two ACIS-I fields) are apparently older than those in the $C$-$field$, and younger than those in the $E$-$field$. The median age for these three groups of stars, indeed, increases toward the east: 1.0 Myear for the stars in the $C$-$field$; 1.4 for the stars in the common region; 2.0 Myears for the stars in the $E$-$field$. This trend contradicts the hypothesis of a large-scale triggering of star formation in M16 by the massive stars in its center, because then the oldest stars should be found in the $C$-$field$. Stars in the $NE$-$field$ are more sparse in the diagram, with a fraction apparently younger than 1Myear. The $J_0$ vs. $(J-K)_0$ diagrams of the same regions, not shown here, share the same properties as the diagrams in Fig. \ref{vvidiag}. Following the extinction map shown in Appendix \ref{extapp}, $A_V$ changes slightly going from the center of the field eastward, with values between $2^m-4^m$. In particular, it is on average smaller in the east direction with respect to the center. This allows us to exclude the hypothesis that this result can be influenced somehow by contamination of the member list (which, however, is not crucial, as explained in Sect. \ref{listsec}). Indeed, if most of the sources in these fields were background sources, their extinction (and then their age) would be underestimated, which mostly affect the sources in the East, which are older that those in the central field in Fig. \ref{vvidiag}. \par

	\begin{figure}[]
	\centering	
	\includegraphics[width=7.5cm]{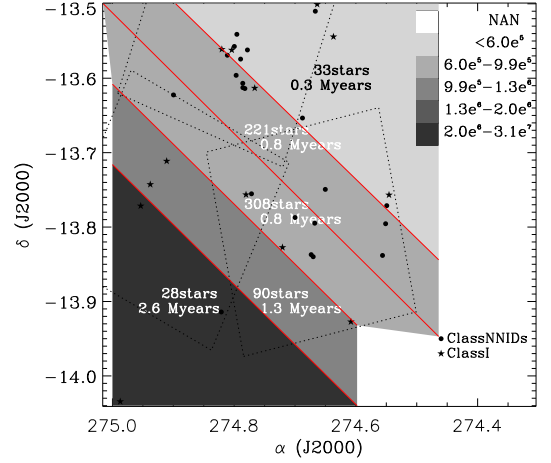}
	\caption{Map of the age of disk-less cluster members. The positions of ClassI/embedded ClassII YSOs are also shown (ClassNNIDs are sources whose IRAC colors allow both classifications). For each region we give the median age of disk-less sources and the number of stars considered for age determination. North is up, East is left.}
	\label{agediag}
	\end{figure}

	In order to investigate the triggering of star formation in M16 by the massive stars associated with NGC~6611 on a large spatial scale, we constructed a map of the age of disk-less cluster members in different regions of M16 dividing the WFI FoV in 25 cells of $1.7\times 1.4$ parsecs. The representative age of each cell was chosen as the median age of all disk-less members falling in that cell. The cell in which most of NGC~6611 falls is 0.8 Myears old, younger than all surrounding cells in every direction but northward. Among the cells with a significant number of stars, the oldest cells (2-3 Myears old) are in the East and South; the youngest (0.1 Myears) is northward from NGC~6611. This grid is not shown here. However, despite an in-outward sequential star formation in M16, which should to be observed in the case of a large-scale triggering of star formation by the massive stars associated to NGC~6611, the grid we obtained suggests an age trend from southeast to northwest. This age gradient is shown in Fig. \ref{agediag}. This map contradicts the hypothesis of a large-scale triggering of star formation by the massive stars of NGC~6611. The oldest disk-less cluster members are in the southeast region (2.6 Myear), while the youngest are in the northwest region (0.3 Myears). Stars in the region of M16 which corresponds to NGC~6611 are 1.3-0.8 Myears old. This trend is confirmed not only by the median values, but also by the ``global offset'' of the whole age distributions toward younger ages going from southwest to northeast. Figure \ref{agediag} also shows the spatial distribution of ClassI/embedded ClassII objects. These stars are sparse, with a group clustered in the northeast, in the center of the NE-field. This group corresponds to the position of the embedded NE cluster (see Sect. \ref{intro}) and the chronology of star-formation we found suggests that this is a young star formation site in M16. \par

%%%%%%%%%%%%%%%%%%%%%%%%%%%%%%%%%%%%%%%%%%%%%%%%%%%%%%%%%%%%%%%%%%%%%%%%%%%%%%%%%%%%%%%%%%%%%%%%%%%%%%%%%%%%%%%%%%%%%%%%%%%%%%%%%%%%%%%%%

\section{Discussion}
\label{discsec}

\subsection{Disk frequency and externally induced photoevaporation}
\label{8.1}

In Sect. \ref{diskfresec} we found an average disk frequency in the $C$-$field$ equal to $36\% \pm 1\%$. In the same sky region, GMD09 evaluated an average disk frequency of $24\% \pm 2\%$, lower than our value. This discrepancy is not surprising, since the new IR photometry is deeper (about $1.5^m$ in $J$) and the new list of members is larger than that found in GMD09, in agreement with the smaller photometric errors in UKIDSS and the higher sensitivity and spatial resolution in UKIDSS/GPS with respect to 2MASS/PSC (allowing a more complete selection of disk-bearing stars). Besides, as explained in the next section, the disk frequency is expected to increase for stars with decreasing masses. In GMD09, the obtained disk frequency in the central region of NGC~6611 was compared with the value estimated by \citealt{Oli05} ($\sim 58\%$), with the conclusion that the discrepancy is due to the deeper infrared observation used in the latter work, and not to an inefficiency of the $Q$ method to select stars with a disk. The new average disk frequency estimated in this work, which is higher than that found in GMD09, supports this hypothesis. \par
	In order to search for a possible relation between disk frequency and stellar mass, we evaluated the mass of the candidate cluster members from the de-reddened $V_0$ vs. $(V-I)_0$ and $J_0$ vs. $(J-K)_0$ diagrams (see Sect. \ref{diskfresec}). We limited this analysis to the $C$-$field$ because NGC~6611 is completely contained in this field. Table \ref{diskfreqtable} shows the average disk frequencies in two bins of stellar mass. A higher frequency has been found for stars with mass between 0.2 and 1 $M_{\odot}$, in agreement with the results obtained in other star-forming regions, as $\eta$ Cha \citep{Mege05}, Upper Sco \citep{Carpe06}, Trumpler~37 \citep{Sici06} and IC348 \citep{La06}. However, these studies evaluated for the population of stars more massive than $1 M_{\odot}$ disk frequencies from few percents to $\sim10\%$, which is significantly lower than our $38 \%$. This difference is in part explained by the different methods used to select disk-less and disk-bearing members. Most of this discrepancy, however, is due to the different age of the clusters. Indeed, all these clusters have an age ranging between 3-9 Myears, older than the members of NGC~6611 (whose median age is lower than 1 Myear). The disk frequency found in our work consequently refers to a younger pre-main sequence population, with a larger number of intermediate- to high-mass stars that have not yet dissipated their disks. \par
With the more complete disk selection of this work, the trend of disk frequency with the incident flux found in GPM07 and GMD09 changed slightly (as shown in the left panel of Fig. \ref{istofig}). In that work the disk frequency decreases monotonically from larger to smaller distances from OB stars. With the new and more reliable list of cluster members, the disk frequency is almost constant in the field, dropping only close to the massive stars, where disks are irradiated by intense UV radiation. Our disk selection is sensitive to stars with an inner disk and those with a population of small grains in their disk. The decrease of the percentage of disks selected by NIR excesses means a dissipation of the whole disks (down to the inner region) and/or of the small ($\sim$ some $\mu m$) grain population. \citet{Io09} discussed several processes which lead to a decrease of the small grains population in photoevaporating disks. \par
	When externally induced photoevaporation occurs, disks are destroyed from outside (where the disk surface density is lower) inward, down to the gravitational radius ($R_g$) where the escape velocity is equal to the sound speed (see \citealt{Holle94}). For instance, in a disk around a 0.2 $M_{\odot}$ star heated up to 1000 K by incident Far-Ultraviolet (FUV) radiation, $R_g$ is at 20 AU. $R_g$ decreases significantly when the disk is irradiated by the Extreme Ultraviolet (EUV) radiation and it is heated up to $\sim$10000 K. The dissipation of the inner disk is also more rapid in EUV regime, occurring in timescales smaller than 1 Myear \citep{Sto99}. This is crucial in our study, because our disk diagnostic is sensitive to the emission from the inner region of the disk and the average age of stars in NGC6611 is $\sim$1 Myear. Extreme ultraviolet radiation dominates the heating of the disks close to OB stars, up to an average distance of about 1 parsec. The left panel of Fig. \ref{istofig} shows that the disk frequency drops close to OB stars, while it is almost constant at larger distance. This is consistent with the observation that in NGC~6611, with an age of $\sim$1 Myear, only disks irradiated by EUV radiation have been significantly dissipated down to the inner region, while FUV radiation in the outer part of the cluster has not affected the disk evolution. Our results agree with those of \citet{Balo07}, who found that in NGC~2244 (a young cluster hosting a O star) the disk frequency drops at a distance smaller than 0.5 parsec from this O star, and with the HST observations in the Trapezium in ONC, where photoevaporating disks have been observed at distance from the O stars smaller than 0.4 parsec. The bin with the highest UV flux in Fig. \ref{istofig} indeed, corresponds to an average distance from massive stars of $\sim$1 parsec. These results go in the direction that only EUV radiation can affect the planetary formation in circumstellar disks, because photoevaporation in the FUV regime could be important only on timescales larger than the lifetime of O stars or than the cluster relaxation time.

	\begin{figure}[]
	\centering	
	\includegraphics[width=9cm]{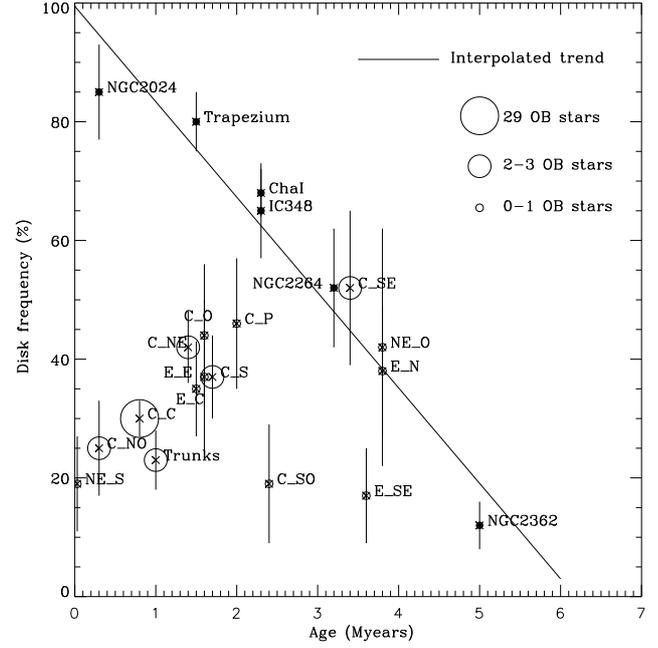}
	\caption{Disk frequencies (in percentage) vs. age for the stellar population of the clusters studied in \citealt{Hai01} (black filled dots) and the regions of the Eagle Nebula defined in Fig. \ref{diskfreqmap} (empty circles). The line marks the age vs. disk frequency relation interpolated from the Haisch data. The size of the empty circles is related to the content of OB stars in the regions, as explained in the figure.}
	\label{regdiag}
	\end{figure}

We also searched for a relation between the observed disk frequency and the median age of cluster members in the regions of the Eagle Nebula defined in Fig. \ref{diskfreqmap}. To this end, we compared the observed disk frequencies with those expected on the basis of the age of the clusters analyzed in \citet{Hai01}. Figure \ref{regdiag} shows the disk frequencies of both the clusters studied in \citealt{Hai01} (filled circles) and the region of the Eagle Nebula defined in Fig. \ref{diskfreqmap} (empty circles). The size of the empty circle is related with the amount of massive stars in the region, as shown in the figure. The disk frequency of these regions can be easily compared with the interpolated {\it age vs. disk frequency} relation, which is marked by the line. Because three of the four regions with the difference between the observed frequency and the corresponding expected frequency (from the interpolated relation) larger than $60\%$ are among those with the largest population of OB stars, a connection between the number of massive stars in the sub-fields and the difference between ``expected'' and observed disk frequencies can be assumed. \par
	Note that the regions with the highest differences are also the youngest. It is possible that in these regions most of the associated YSOs with disk are very embedded in the cluster medium, which makes their detection hard and partially accounts for the observed differences. The only field in which this hypothesis does not work is the $C\_C field$, which corresponds to the cavity cleared by NGC~6611 stars (and among the region with lower extinction, see Appendix \ref{extapp}). In $C\_C field$ the difference is more likely due to the high number of massive stars. In conclusion, Fig. \ref{regdiag} shows a larger difference between the observed and the ``expected'' disk frequency for those regions that are younger average and for those with more massive stars. \par

\subsection{Efficiency of disk photoevaporation for different masses of the central star}
\label{8.2}
	
	The right panel of Fig. \ref{istofig} shows how the disk frequency for low- and intermediate- to high-mass stars varies with the incident flux emitted by OB stars. This analysis is performed to study the relative role of two different mechanisms for disk dissipation which can take place in young massive clusters. For a unperturbed evolution, the timescale for disk dissipation strongly depends on the mass of the central star. Even if there is not yet an universal scaling law which gives the time of dissipation of the inner disk versus the mass of the central star, high-mass stars emit more energetic flux and more intense stellar winds than low-mass stars, therefore they erode their inner disks faster. Indeed, the disk frequencies for high-mass stars decline faster than those evaluated on the entire cluster populations (which declines in $\sim10$Myears, \citealt{Hai01}). On the other hand, for disks photoevaporated by close massive stars, specific simulations predict a faster dissipation of disks in low-mass stars with respect to high-mass stars. For instance, \citet{Ada04} have shown that the intensity of the flux of photoevaporating gas from disks with a central star of $0.25 M_{\odot}$ is less intense (by one order of magnitude) than those with a star of $1 M_{\odot}$. This is due to the weaker gravitational field produced by less massive stars, which also results in a larger gravitational radius.  \par
In the right panel of Fig. \ref{istofig}, the difference between the frequencies for high- and low-mass stars is almost independent from the intensity of the incident flux. This suggests that in NGC~6611 the disk dissipation timescales due to photoevaporation induced by OB stars do not strongly depend on the mass of the central stars. This can be explained by the age of the cluster ($\sim$1 Myear). Indeed, following the simulations of \citet{Ada04} and using the typical mass-loss rates which they evaluated, the total mass lost in 1 Myear from two disks around a 1 $M_{\odot}$ and 0.5 $M_{\odot}$ star, respectively, differs by less than $\sim 0.01 M_{\odot}$. This low value cannot produce observable differences in NIR excesses. \par

\subsection{Large-scale star-formation triggering induced by OB stars}\par
In Sect. \ref{agesec} we tested the hypothesis of a possible large-scale triggering of star-formation in the whole M16 by the massive stars in NGC~6611, searching evidence that younger objects are concentrated in the outer regions of the Eagle Nebula. The age trend shown in Fig. \ref{agediag} negates this hypothesis: only in the northward direction the disk-less stars associated with the nebula are younger than those in the central region, while in the other direction older objects have been found. Instead, there is a gradient trend in the SE-NW direction, with a difference of 2.3 Myears between the two extremes (see Fig. \ref{agediag}). Hence, the picture which arises from this is that the star-formation events in the southeast region of M16 are older (of 1-2 Myears) than those in NGC~6611 (central region). Only the star formation in the North and West (i.e. the region around the massive star in formation and the SFO30 cloud) may have been triggered by the radiation from the massive stars of NGC~6611 in recent time (less than 1 Myear), as expected by a small-scale triggering phenomenon. The star formation in the embedded cluster in the North-East occurred in the last 0.5-0.3 Myears.\par
	If this sequence of star formation is correct, something has given the initial impulse for the star formation in the South-East about 3 million years ago. We tentatively identified this event in the incidence on M16 of a giant molecular shell discovered by \citet{Mori02}. Following these authors, this shell is about $170 \times 250$ parsecs wide, and was created about 6 Myears ago likely by supernova explosions. This shell reached the Eagle Nebula in the southern direction approximately 2-3 Myears ago. \par

	\begin{figure}[]
	\centering	
	\includegraphics[width=9cm]{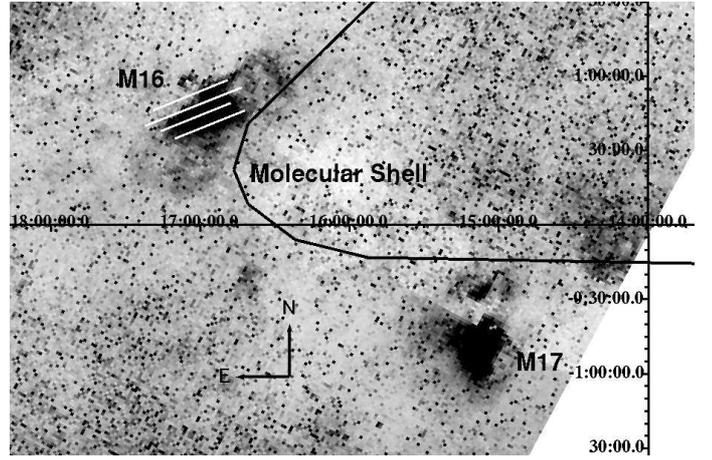}
	\caption{DSS-I image of about $4^{\circ} \times 3^{\circ}$ of the sky region between M16 and M17. The axis of the galactic coordinates, the limit of the molecular shell (black curve), and the limits of the regions in Fig. \ref{agediag} (white lines) are shown.}
	\label{shellfig}
	\end{figure}

Figure \ref{shellfig} shows a DSS-I $4^{\circ} \times 3^{\circ}$ image of the region of the galactic plane with the Eagle Nebula (M16) and the Omega Nebula (M17). The limits of the molecular super-shell discovered by \citealt{Mori02} (the black curve) and the limits of the diagonal regions in Fig. \ref{agediag} (the white lines) are also shown, with the oldest region of Fig. \ref{agediag} closer to the shell. Both the direction and the time at which the molecular shell encountered the Eagle Nebula are consistent with the sequence of star-formation suggested by this work. Moreover, both the star-formation events in the southeast region of M16 and the oldest events in the Omega Nebula (M17) can be dated to approximately 3 million years ago (see \citealp{Povi09}), which supports a connection between the stars-formation events in the two star-forming regions.

%%%%%%%%%%%%%%%%%%%%%%%%%%%%%%%%%%%%%%%%%%%%%%%%%%%%%%%%%%%%%%%%%%%%%%%%%%%%%%%%%%%%%%%%%%%%%%%%%%%%%%%%%%%%%%%%%%%%%%%%%%%%%%%%%%%%%%%%%

\section{Conclusions}
\label{thatsallfolk}

We studied the chronology of star formation in the Eagle Nebula (M16) and the spatial variation of disk frequency to test the effects of OB stars on the parental cloud and on the nearby population of young stars. We analyzed optical, infrared, and X-ray data of this region. X-ray data have been obtained from the ACIS archive (one observation centered on NGC~6611 in the center of M16) and from two new observations centered on the pillar structure called ``ColumnV'' and on a recently discovered embedded cluster (in the East and northeast regions of M16, respectively). We detected 1755 X-ray sources inside these three regions. Infrared data in $JHK$ bands of the whole region have been obtained from the UKIDSS/GPS catalog. Optical $BVI$ and Spitzer/IRAC data of M16, used in this work, have been previously analyzed in other publications. With all these data, we compiled a multiband catalog of this region. \par
	We selected the candidate stars associated with M16 using two criteria: infrared excesses for disk-bearing sources and X-ray emission for disk-less stars. Infrared excesses have been detected with two different diagnostics: the IRAC color-color diagram and optical-infrared color indices defined in order to be independent from the extinction. We selected a total of 834 candidate disk-bearing stars. The list of young stars associated with M16 has been completed with 1103 candidate disk-less members, for a total of 1937 sources.\par
We have found an average disk frequency of $36\% \pm 1\%$ in the central region of the nebula, which corresponds to NGC~6611. We also found a weak dependence of the disk frequency on stellar mass: the disk frequency is slightly higher for stars less massive than $1M_{\odot}$ with respect to more massive stars. For stars more massive than $1\,M_{\odot}$, the frequency is equal to $32\% \pm 4\%$, higher than the values observed in older star-forming regions, likely because in 1 Myear (which is the median age of NGC~6611) a significant number of high-mass stars have not yet dissipated their disks. \par
	Disk frequency drops at small distance (d$\leq 1pc$) from the massive stars of NGC~6611, where disks are irradiated by intense EUV fluxes. The variation of the disk frequency with the intensity of incident UV flux does not depend on the mass of the central stars. These results indicate that in $\sim$1 Myear photoevaporation induced externally by EUV radiation has completely dissipated only disks close ($d\leq 1$ parsec) to OB stars, without observable effects at larger distance (where photoevaporation is driven by FUV photons) and without an evident dependence of the efficiency of photoevaporation on the mass of the central stars. This study supports the hypothesis that only EUV radiation from OB stars may have an effect on the planet formation inside circumstellar disks.\par
We studied the history of star formation in the nebula, finding that YSOs in NGC~6611 are on average younger than those in the regions of M16 in the East and South, and older than those in the North. We derived a chronology of star formation in the nebula from southeast (about 2.6 Myears ago) to northwest (less than 1 Myear ago). This chronology of star formation does not support the hypothesis that OB stars in NGC~6611 have triggered the formation of new stars in the whole nebula up to a distance of $\sim$10 parsec. We explained this chronology with a star formation process induced externally, and we tentatively identified a giant molecular shell, which reached M16 about 2-3 million years ago, starting the star formation, and which could connect the Eagle Nebula with the Omega Nebula. \par

%%%%%%%%%%%%%%%%%%%%%%%%%%%%%%%%%%%%%%%%%%%%%%%%%%%%%%%%%%%%%%%%%%%%%%%%%%%%%%%%%%%%%%%%%%%%%%%%%%%%%%%%%%%%%%%%%%%%%%%%%%%%%%%%%%%%%%%%%
\begin{acknowledgements}
This work is based in part on data obtained as part of the UKIRT Infrared Deep Sky Survey and with observations performed with CHANDRA/ACIS, public data obtained with WFI@ESO, 2MASS Point Source Catalog and GLIMPSE survey with Spitzer/IRAC. Support for this work has been provided by the CONSTELLATION grant YA 2007 and the contract PRIN-INAF (P.I.: Lanza) and by the ASI/INAF contract I/088/06/0. The author thanks Jeremy Drake and Francesco Damiani for their important suggestions. We also thank the anonymous referee, whose review helped us to improve our paper and make our final results much more robust.
\end{acknowledgements}

\appendix
\onecolumn
\section{YSO population in selected regions of M16}
\label{popusec}

\begin{figure}[]
	\centering	
	\includegraphics[width=9cm]{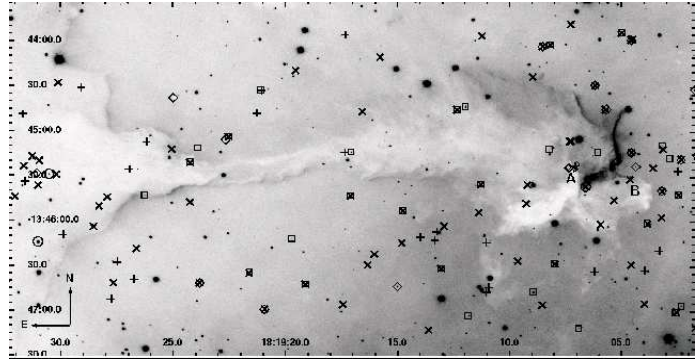}
	\caption{VIMOS@ESO image of the ColumnV in R band. Symbols are described in Table \ref{symtable}.}
	\label{colufig}
	\end{figure}

\begin{figure}[]
	\centering	
	\includegraphics[width=9cm]{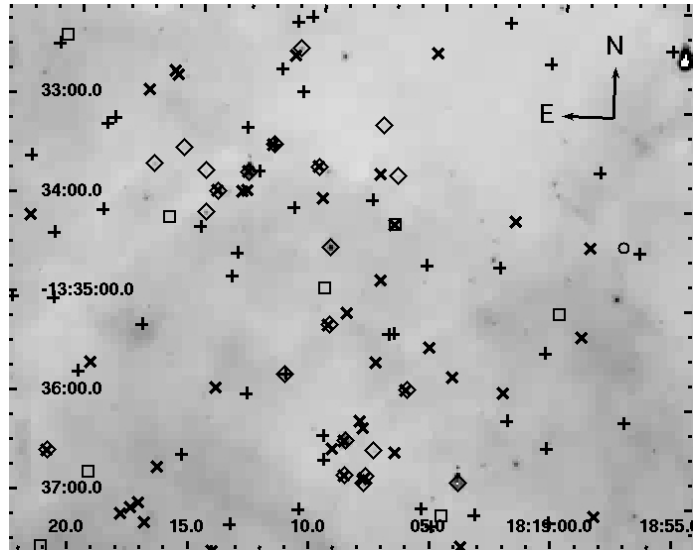}
	\caption{IRAC [8.0] image of the embedded cluster in the North-East. Symbols are described in Table \ref{symtable}.}
	\label{clunefig}
	\end{figure}

\begin{figure}[]
	\centering	
	\includegraphics[width=9cm]{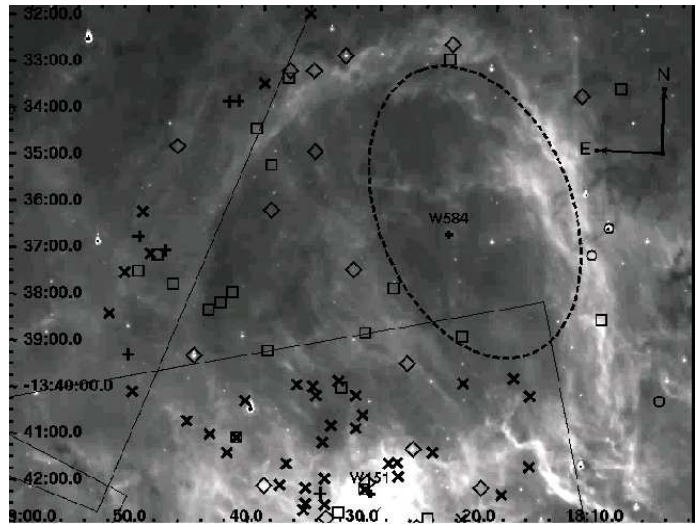}
	\caption{IRAC [8.0] image of the region in the northwest of the Eagle Nebula. Symbols are described in Table \ref{symtable}. Boxes are the ACIS-I FoVs, while the dashed line marks the sky region inside the cavity, close to W584, where no disk-bearing stars have been selected.}
	\label{cavifig}
	\end{figure}

In this appendix we briefly describe the candidate members of M16 associated with the ColumnV, with the embedded cluster in the northeast and with the cavity at northwest. In the images of these regions, Figs. \ref{colufig}, \ref{clunefig}, and \ref{cavifig}, the candidate members are marked with the symbols described in Table \ref{symtable}.

	\begin{table}[]
	\centering
	\caption {Symbols used to mark the stars in Figs. \ref{colufig}, \ref{clunefig}, and \ref{cavifig}}
	\vspace{0.5cm}
	\begin{tabular}{cc}
	\hline
	\hline
	Type of star & Symbol  \\
	\hline
	X-ray sources with a stellar counterpart	&X	\\
	X-ray sources without any counterpart		&pluses	\\
	Stars with disk added from 2MASS member list	&circles	\\
	Stars with excesses only in UKIDSS bands	&squares	\\
	Stars with excesses in IRAC bands		&diamonds	\\
	\hline
	\hline
	\multicolumn{2}{l} {} 
	\end{tabular}
	\label{symtable}
	\end{table}

Figure \ref{colufig} shows a 10$\,$sec image in $R$ band of the ColumnV, obtained with VIMOS@ESO on 2009-03-24 (obs. ID: 083.C-0837, PI: Guarcello) as part of the VIMOS $pre-imaging$ of the Eagle Nebula. The sky region in this image is $7.8^{\prime}\times 3.6^{\prime}$ wide (corresponding to $3.7\times 1.7$ parsec), and it is $7^{\prime}$ (3.3 parsec) eastward from NGC~6611. In this region are 97 X-ray sources (among which 23 have also infrared excesses, 22 do not have any stellar counterpart, 52 are candidate disk-less YSOs), 52 candidate disk-bearing stars (31 with only UKIDSS excesses; 18 with IRAC excesses; 3 added from the 2MASS list of members), for a total of 104 candidate members and an average disk frequency equal to $50\% \pm 8\%$. In the various color-color and color-magnitude diagrams, the positions of all sources in Fig. \ref{colufig} indicate that they are on average affected by the typical visual extinction of the optical sources associated with the nebula ($A_V \sim 2.5^m-3^m$), and that they are on average 1-1.5 Myears old. Following the star-formation chronology across the Eagle Nebula shown in Fig. \ref{agediag}, it can be assumed that the ColumnV and the Elephant Trunks have been formed almost simultaneously in the last 1-1.5 million years. Indeed, the few sources younger than 1 Myears are those in front of the cap of the pillars. The massive stars closest to the ColumnV are W541 ($\alpha=$18:19:19.123, $\delta=$-13:43:52.32), a binary \citep{Marta08} B2.5 star \citep{Hille93} $1.3^{\prime}$ northward, and W472 ($\alpha=$18:19:04.709, $\delta=$-13:44:44.52), a B3 star \citep{Hille93}, in correspondence of the cap of the pillar. The source W541 may be responsible for the erosion of the center of the pillars, while W472 may be the ionizing source which heats the cap of the pillar and may have induced the more recent star-formation events. In Fig. \ref{colufig}, the letter ``A'' marks the position of four water masers identified by \citet{Hea04}. One embedded candidate ClassI YSO, with very red $[3.6]-[4.5]$ color, can be associated with these water masers (see also \citealp{Inde07}). We also identified an X-ray source without stellar counterpart, $1.6^{\prime \prime}-2^{\prime \prime}$ away from the water masers. This source could also correspond to a very embedded YSO associated with the water masers, also because of its hard spectrum (its median photon energy is equal to 3.5 keV). The region in front of the pillar is richly populated: 7 disk-less X-ray sources and 14 candidate stars with disk (among which 8 X-ray sources). The point ``B'' in Fig. \ref{colufig} marks an ionized knot candidate to be a Herbig-Haro object \citep{Mea86}. A soft (median energy equal to 1.9 keV) X-ray source and a faint stellar counterpart is $1^{\prime \prime}$ northward from the knot. \par

Figure \ref{clunefig} shows the IRAC [8.0] image of the sky region corresponding to the embedded cluster in the North-East ($7.3^{\prime} \times 5.3^{\prime}$ wide, corresponding to $3.4 \times 2.5$ parsec). This region includes 59 X-ray sources (among which 24 have no stellar counterpart, 9 have infrared excesses and 26 are disk-less members), and 25 candidate stars with disk (21 with excesses in IRAC bands, 4 only in UKIDSS bands). Both the extinction map, the color-magnitude diagrams and the age trend (Figg. \ref{absjhhkfig} and \ref{agediag}) show that this region is populated by very young (age $\leq 1$ Myear) and embedded sources ($4^m\leq A_V\leq 20^m$). Disk-bearing stars in this region are heavily extincted, so they lack optical counterpart (except one star with faint $I$ emission) and they are very red in UKIDSS bands ($J-K \geq 3^m$). Also in the IRAC color-color plane, objects in this region can be classified as very embedded ClassII/ClassI sources (14 stars), with also 4 ClassII sources with moderate reddening and 1 ClassI YSOs. A very hard X-ray source is associated with this ClassI YSO, with 31 net counts and a median energy of 5.5 keV. According to the high extinction, in this region the X-ray sources with infrared counterpart have very red colors and, on average, hard spectra, with typical median energies of 3 keV. \par

Figure \ref{cavifig} shows the sky region in the northwest of NGC~6611 with the side of approximately $10^{\prime}$ (4.8 parsecs) which has not been observed in X-ray (the solid lines in Fig. \ref{cavifig} indicate part of the ACIS-I FoVs). In this region, M16 has a cavity-like structure, with the O9V W584 star near the center of this cavity, at $\sim0.7$ parsec from the well defined NW border of the cavity. This kind of structure could have been created by the intense wind emitted by this massive star. As shown in Fig. \ref{cavifig}, no star with disk falls close to W584, while 20 candidate disk-bearing members are selected in every direction along the border of the cavity. The formation of these stars could have been induced by the compression of the nebula by W584. Since it is more likely that the nebula has been swept away continuously from the position of W584 outward, formation of new stars should be induced also inside the cavity, and not only in its present borders. Then, it is possible to suppose that W584 has not only created the cavity, but also induced a fast dissipation of the circumstellar disks in these young stars. An X-ray follow-up could be useful to select candidate disk-less YSOs inside the cavity, to clarify this point. 

\newpage
\section{Extinction map of the Eagle Nebula}
\label{extapp}

	\begin{figure}[]
	\centering	
	\includegraphics[width=9cm]{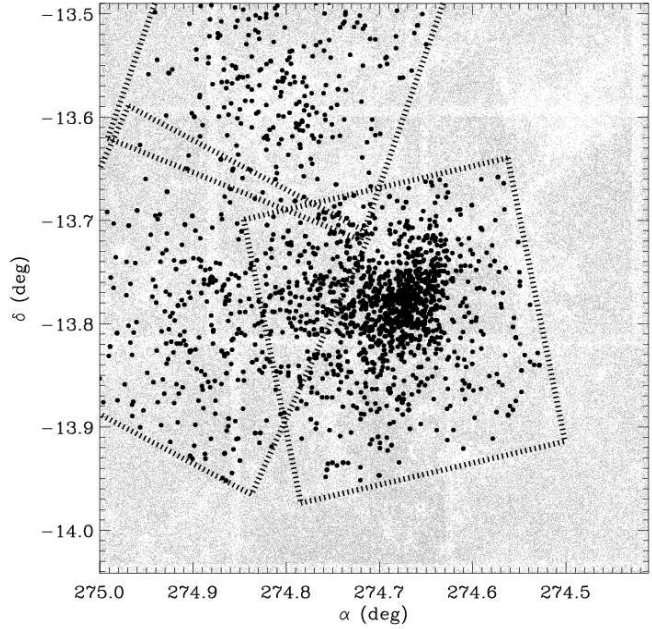}
	\caption{Spatial distribution of the UKIDSS sources (gray points) and of the X-ray sources (black dots). Dashed boxes mark the ACIS-I fields. North is up, East is on the left}
	\label{ukxspadis}
	\end{figure}

In Sects. \ref{diskfresec} and \ref{agesec} we used the individual extinction of cluster members in order to derive both stellar masses and ages. It was not possible to adopt an average extinction for all stars associated with M16 because the density of the cloud varies significantly across the field, as it is evident from the spatial distribution of UKIDSS sources (marked with gray points) shown in Fig. \ref{ukxspadis}. Then, stellar extinction values have been evaluated by means of a map of visual extinction $A_V$ toward the Eagle Nebula. Figure \ref{ukxspadis} also shows the spatial variation of X-ray sources (black dots). Because they are dominated by cluster members, X-ray sources are clustered on NGC~6611 (in the center of the $C$-$field$) with a smooth decrease of source density eastward. The region in the $NE$-$field$, where falls a group of X-ray sources in a region of high nebulosity, corresponds to the site of the northeast embedded cluster (see Appendix \ref{popusec}). \par 
	Usually, using photometric data to derive the extinction map of a given region, it is necessary to identify a sample of stars whose intrinsic colors are accurately known. In this way, the interstellar reddening can be evaluated as the difference between the observed and the intrinsic colors of these stars. Usually giant stars, whose average infrared colors have been evaluated by several authors (i.e. \citealp{Besse88}), are used to do this. In our case, giant stars dominate the sample of UKIDSS sources without optical counterpart, but their extinctions are dominated by the unknown distribution of background interstellar medium in our Galaxy from M16 up to very large distances ($\sim 10 Kpc$, as suggested by \citealp{Luca08} and by the $J-H$ vs. $H-K$ diagram for stars in the WFI FoV), and not by the material associated with M16. \par
It is possible to obtain an accurate estimate of the visual extinction toward the Eagle Nebula by considering the disk-less X-ray sources with NIR emission, because they are mostly PMS stars associated with the nebula and their emission is not affected by the background interstellar medium. Typically, T-Tauri stars with an age of 1 Myear and $H-K\ge0.2^m$ are of K-M spectral type \citep{Mey97}. Their intrinsic $(J-H)_0$ ranges from 0.5$^m$ to 0.7$^m$, with a median value of 0.66$^m$ \citep{Sie00}. Assuming this value, it is possible to estimate the individual $A_V$ from the relation $E_{J-H}=(J-H)_{observed}-(J-H)_0$ and $E_{J-H}=0.107\times A_V$. In order to obtain reliable results, it is mandatory to exclude candidate disk-bearing stars, whose infrared colors are not photospheric. Our selection of disk-bearing stars is discussed in detail in Sect. \ref{disksec}. Starting from the individual $A_V$, we derived the extinction map by dividing the WFI FoV  in 100 squared cells and we assigned to each cell the median value of the extinction values of all selected stars falling in that cell. \par

	\begin{figure}[]
	\centering	
	\includegraphics[width=9cm]{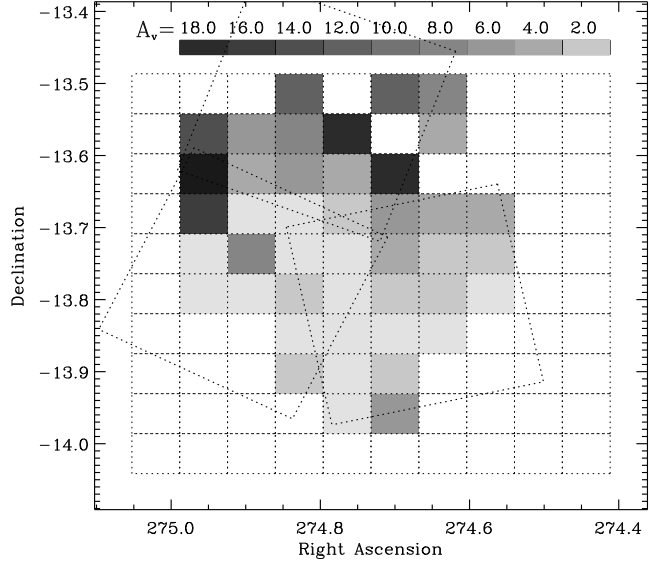}
	\caption{$A_V$ map in the WFI FoV obtained from candidate disk-less members of M16. The dotted boxes are the ACIS-I FoVs analyzed in this work.  $A_V$ cannot be computed in the white boxes because of the lack of disk-less members with adequate quality photometry in $J$ and $H$ bands. North is up, East is on the left.}
	\label{absjhhkfig}
	\end{figure}

Figure \ref{absjhhkfig} shows the M16 extinction map obtained with the technique described above. The extinction map shows an increase of visual extinction from the central cavity, corresponding to NGC~6611, northward. $A_V$ peaks in correspondence of some structures of the nebula described in Sect. \ref{m16}, as the cluster in the North-East. The increase of extinction northward has been already found by \citet{Beli99}. \par
	With the standard {\it hydrogen column} ($N_H$) vs. {\it visual extinction} relation \citep{Prede95}, the difference of the average extinction between the central and the outer regions of the nebula implies an average hydrogen column density which ranges from $4.7\times10^{21}$ atoms/cm$^2$ in the center to $3.2\times10^{22}$ atoms/cm$^2$ in the denser regions. \par

 	\begin{figure}[]
	\centering	
	\includegraphics[width=7cm]{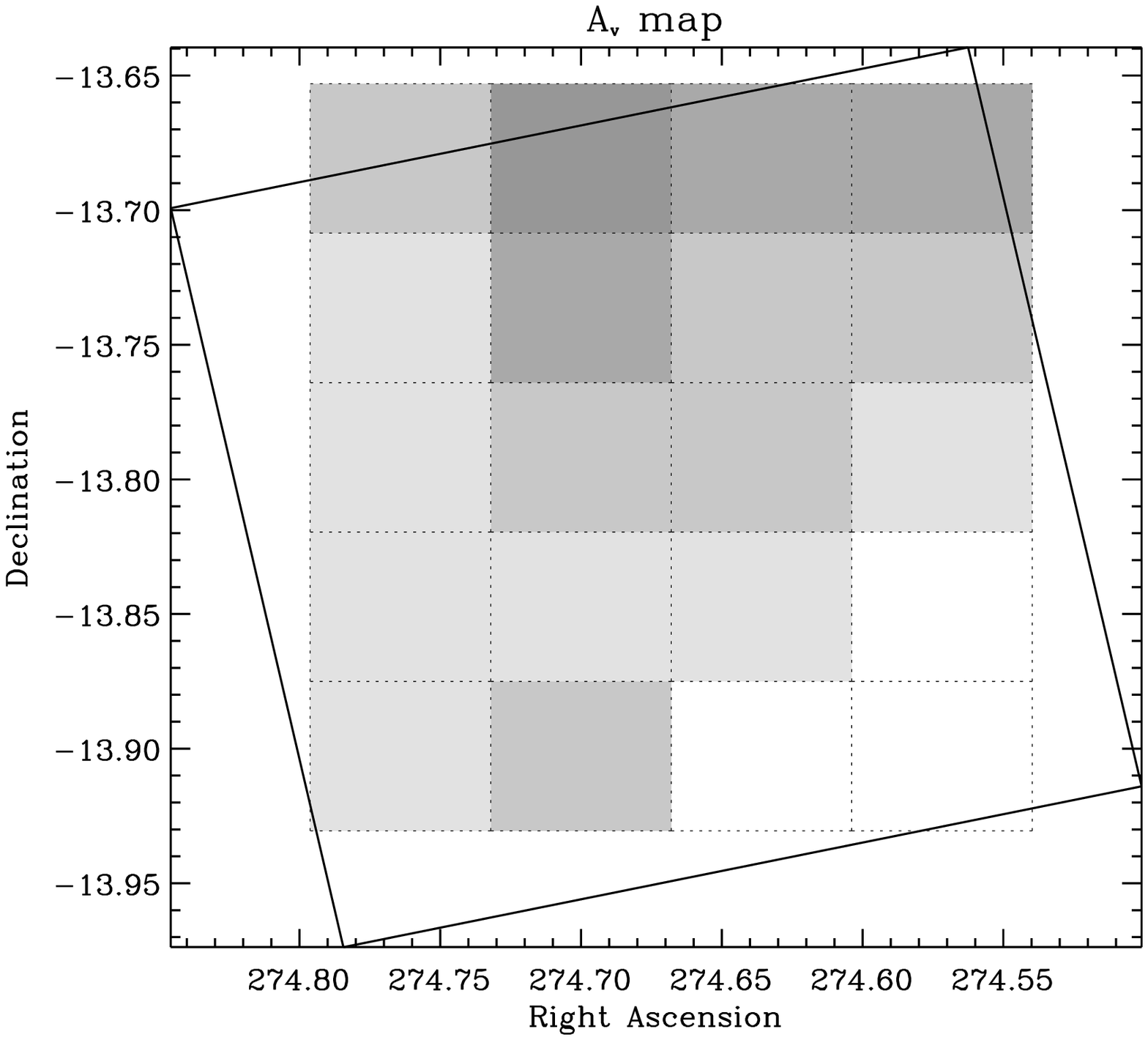}
	\includegraphics[width=7cm]{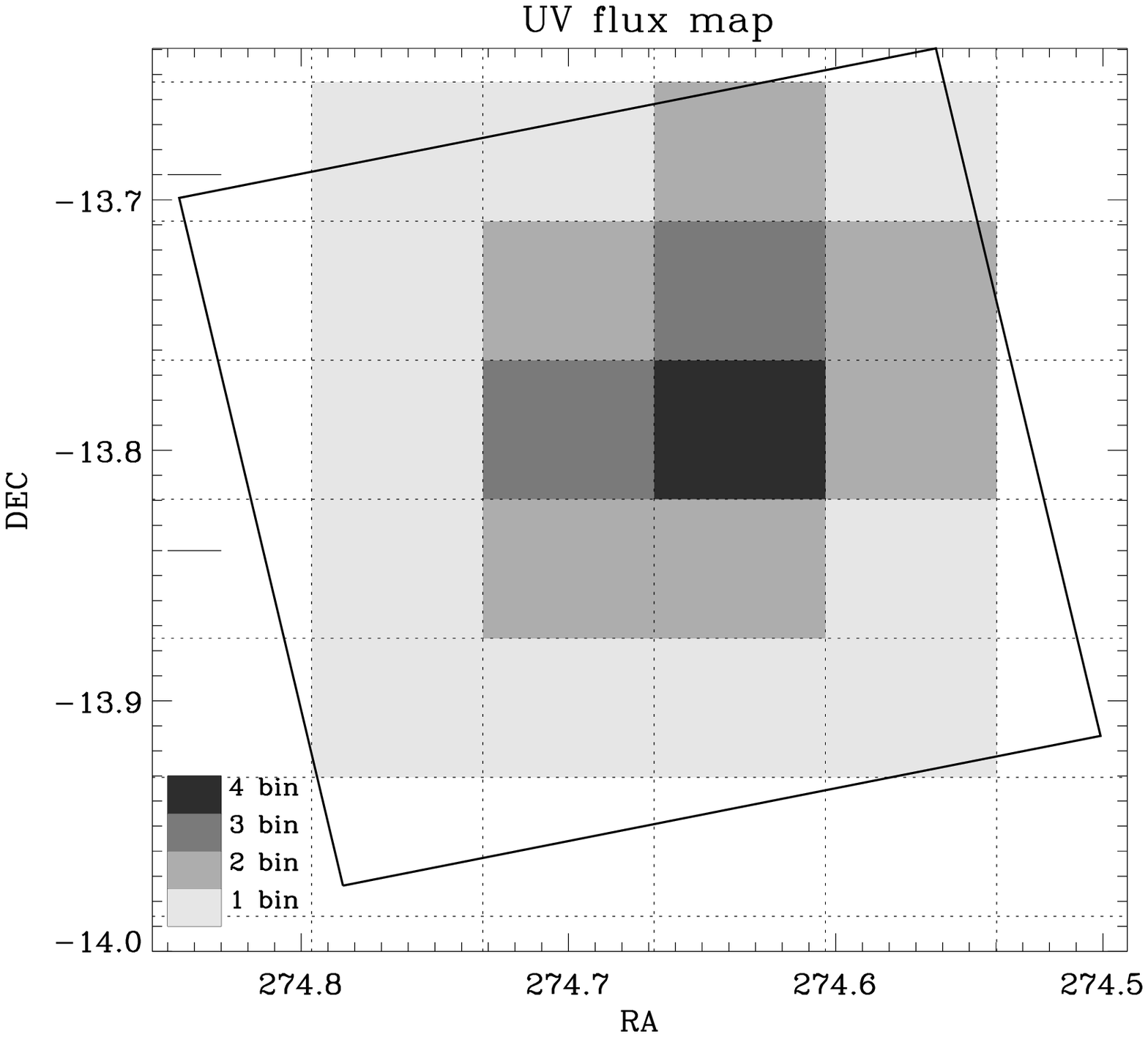}
	\caption{Left panel: map of the visual extinction of the central region of M16, corresponding to NGC~6611. The gray scale is the same of that used in Fig. \ref{absjhhkfig}. Right panel: map of the intensity of the UV field across NGC~6611. The gray scale relates the median intensity of the UV field to the bins in the left panel of Fig. \ref{istofig}. In both panels the inclined box delimits the ACIS-I FoV. North is up, East is on the left.}
	\label{uv_avfig}
	\end{figure}

In order to verify that the drop of the disk frequency close to the massive stars (see Fig. \ref{istofig}) is not due to a larger numbers of background contaminants selected as disk-bearing cluster members in the outer region of NGC~6611, resulting in higher disk frequency values, we compare the visual extinction inside the $C-field$, where NGC~6611 lies, with the spatial variation of the UV field using Fig. \ref{uv_avfig}. The left panel shows the $A_V$ map, which is a magnification of the central part of Fig. \ref{absjhhkfig}. The gray-scale is also the same. The right map has been created evaluating the UV field incident on the position corresponding to the center of each bin, taking into account the flux emitted by all massive stars of NGC~6611. The gray scale relates the intensity value to the bins in the histogram in Fig. \ref{istofig}. Comparing both images, no correlation between the intensity of the UV field and the visual extinction, i.e.: with the density of the cloud, arises. This leads to the conclusion that our result about the spatial variation of disk frequency is not affected by background contamination. 

\newpage
\section{Cross-correlation procedure}
\label{matchapp}

	Here we describe the procedure we adopted to cross-correlate all catalogs in the field of M16 used in this work. The match is performed by identifying the sources belonging to different catalogs with a separation smaller than a given matching radius. The used matching radii have been chosen as the highest values at which the spurious matches are negligible with respect to the real cross-identifications. The expected number of spurious matches at a given matching radius have been evaluated with the method described in \citet{Dami06}. \par
In order to build the joined catalog of X-ray sources, we noted that the overlap region among $E$-$field$ and $NE$-$field$ is very small (Fig. \ref{ukxspadis}), with only few sources found in both fields. Given the extent of the overlapping region between the $C$-$field$ and $E$-$field$, we found the common sources by matching the corresponding sources list, adopting a matching radius equal to $1^{\prime\prime}$ for sources with off-axis smaller than $7^{\prime}$ and $2^{\prime\prime}$ otherwise (considering the highest value of the source off-axis angle). A total of 76 sources was found in both fields, with no multiple matches. To match the optical and the infrared catalogs we used a matching radius of $0.3^{\prime\prime}$. \par
	Because in the ACIS-I field the PSF degrades at large off-axis, to match optical+infrared and X-ray catalogs it was mandatory to define a matching radius which depends on the sources off-axis ($\theta$) in the X-ray images. Because the uncertainty on source positions grows with the off-axis, as a consequence of the spreading of the PSF, we defined the matching radius (in arcseconds) as

\begin{equation}
r_{match}=A\times f(\theta),
\label{rmatcheq}
\end{equation}

	where $f(\theta)$ is an interpolated relation for the source's off-axis and the position error derived by PWDetect.  Equation \ref{rmatcheq} holds for  off-axis$\geq 3^\prime$, because for smaller off-axis the position errors do not vary with $\theta$. To find the value of $A$, we should compare the real and spurious matches distributions. Because the spatial distribution of the X-ray sources in not even, we were not able to use the method of \citet{Dami06} to evaluated spurious matches (which is based on the hypothesis of catalogs with an almost uniform spatial distribution). To see how the number of casual identifications increases with the matching radius, we produced 40 random X-ray catalogs by rigid translations of the X-ray coordinates (from 1$^{\prime}$ to 4$^{\prime}$ in each direction). Then, we matched with different $A$ the optical-infrared catalog with these random X-ray catalogs, in order to find the distribution of casual coincidences at various $A$. We set $A=1$ because this is the highest value at which the spurious matches are negligible with respect to the real ones. For an off-axis smaller than $3^{\prime}$ we set the matching radius at 0$^{\prime\prime}$.5. The number of multiple matches in our final catalog is reasonably low (860, with 88 produced by matching the X-ray catalog). \par

\twocolumn 
\addcontentsline{toc}{section}{\bf Bibliografia}
\bibliographystyle{apj}
\bibliography{biblio}

\end{document}